\documentclass[aps,prd,amsmath,amssymb,eqsecnum,nofootinbib]{revtex4}

\usepackage{color}
\usepackage{graphicx}       
\usepackage{graphics}

\newcommand{\rc}{\rho}

\newcommand{\sech}{\text{sech}}
\newcommand{\Gret}{G_{\text{ret}}}

\begin{document}

\title{Pad\'{e} Approximants of the Green Function in Spherically Symmetric Spacetimes}
\author{Marc Casals}
\email{marc.casals@dcu.ie}
\affiliation{CENTRA, Instituto Superior T\'{e}cnico, Lisbon, Portugal \\ School of Mathematical Sciences, Dublin City University, Glasnevin, Dublin 9, Ireland}
\author{Sam Dolan}
\email{sam.dolan@ucd.ie}
\author{Adrian C. Ottewill}
\email{adrian.ottewill@ucd.ie}
\author{Barry Wardell}
\email{barry.wardell@ucd.ie}
\affiliation{Complex and Adaptive Systems Laboratory and School of Mathematical Sciences, University College Dublin, Belfield, Dublin 4, Ireland}

\date{\today}

\begin{abstract}
We investigate the scalar Green function for spherically symmetric spacetimes expressed as a coordinate series expansion in the separation of the points. We calculate the series expansion of the function $V(x,x')$ appearing in the Hadamard parametrix of the scalar Green function to very high order. This expansion is then used to investigate the convergence properties of the series and to estimate its radius of convergence. Using the method of Pad\'{e} approximants, we show that the series can be extended beyond its radius of convergence to within a short distance of the normal neighborhood boundary.
\end{abstract}
\maketitle

\section{Introduction} \label{sec:intro}
Quasi-local series expansions -- expansions in the separation of two points $x$ and $x'$ -- are a frequently used tool for calculations of fields on curved spacetimes. Often, as a final step in the calculation, the coincidence limit, $x' \rightarrow x$, is taken. In these cases, the precise convergence properties of the series is of little interest. However, there are cases where we would like the points to remain separated \cite{Casals:Dolan:Ottewill:Wardell:2009,Quinn:Wald:1997,Quinn:2000,Mino:Sasaki:Tanaka:1996,DeWitt:1960}. In particular, we are motivated by the calculation of the \emph{quasilocal} (QL) contribution to the scalar self-force \cite{Ottewill:Wardell:2008,Ottewill:Wardell:2009}  (for a review, see \cite{Poisson:2003,Detweiler:2005}) on a scalar particle,
\begin{equation}
\label{eq:QLSFInt}
f^{a}_\mathrm{QL} (z(\tau))= \lim_{\epsilon \rightarrow 0} q^2 \int_{\tau - \Delta \tau}^{\tau-\epsilon} \nabla^{a} G_{ret} \left( z(\tau),z(\tau') \right) d\tau ',
\end{equation}
where $z(\tau$) describes the worldline of the particle. This requires a quasi-local expansion of the retarded Green function, $G_{ret}(x,x')$, which is a solution of the scalar wave equation with point source, 
\begin{equation}
 (\square - \xi R)\Gret(x, x^{\prime}) = -4 \pi \frac{\delta^4(x^{\mu} - x^{\mu \prime})}{\sqrt{-g}},
 \label{eq:gf-waveeq}
\end{equation}
where $\xi$ is the curvature coupling constant and $R$ is the Ricci scalar. As expression \eqref{eq:QLSFInt} requires the Green function for the points separated up to an amount $\Delta\tau$ along a world-line, it begs the question: how large can the separation of the points be before the series expansion is no longer a valid representation of the Green function?

To the authors' knowledge, this question has not yet been quantitatively answered. It is well known that the Hadamard parametrix for the Green function (upon which quasi-local calculations are based) is valid provided $x$ and $x'$ lie within a \emph{normal neighborhood}\footnote{More precisely, the Hadamard parametrix requires that $x$ and $x'$ lie within a \emph{causal domain} -- a \emph{convex normal neighbourhood} with causality condition attached. This effectively requires that $x$ and $x'$ be connected by a unique non-spacelike geodesic which stays within the causal domain. However, as we expect the term \emph{normal neighbourhood} to be more familiar to the reader, we will use it throughout this paper, with implied assumptions of convexity and a causality condition.\label{def:causal domain}} \cite{Friedlander}. However, this does not necessarily guarantee that a series representation will be convergent everywhere within this normal neighborhood. In fact, we will show that the series is only convergent within a smaller region, the size of which is given by the \emph{circle of convergence} of the series. However, this does not preclude the use of the quasi-local expansion to calculate the Green function \emph{outside} the circle of convergence (but within the normal neighborhood). As we will show, Pad\'{e} resummation techniques, which have been extremely successful in other areas \cite{Damour:Iyer:Sathyaprakash,Porter:Sathyaprakash}, are also effective in extending the series beyond its circle of convergence.

In this paper, we will focus in particular on calculating the Green function for two spherically symmetric spacetimes: Schwarzschild and Nariai. The Nariai spacetime~\cite{Nariai:1950,Nariai:1951} arises naturally from efforts to consider a simplified version of Schwarzschild \cite{Casals:Dolan:Ottewill:Wardell:2009,Cardoso:Lemos:2003,Zerbini:Vanzo:2004}. It retains some of the key features of Schwarzschild (such as the presence of an unstable photon orbit and a similar effective radial potential
which diminishes exponentially on one side), but frequently yields more straightforward calculations. This makes it an ideal testing ground for new methods which are later to be applied to the more complicated Schwarzschild case. In the present work, we will use the line element of the static region of the Nariai spacetime (with cosmological constant $\Lambda=1$ and Ricci scalar $R=4$) in the form
\begin{equation} 
 ds^2 = -(1-\rc^2) dt^2 + (1-\rc^2)^{-1} d\rc^2 + d\Omega_2^2, \quad \quad \quad d\Omega_2^2=d\theta^2+\sin^2\theta d\phi^2   \label{eq:Nariai-le}.
\end{equation}
where $\rc\in (-1,+1),t\in (-\infty,+\infty),\theta\in [0,\pi],\phi\in [0,2\pi)$.
In this form, it yields a wave equation with potential which is seen to closely resemble that of the Schwarzschild metric,
\begin{equation} 
 ds^2 = -\left(1-\frac{2M}{r}\right) dt^2 + \left(1-\frac{2M}{r}\right)^{-1} dr^2 + r^2 d\Omega^2_2,   \label{eq:Schw-le}
\end{equation}

In Sec.~\ref{sec:WKB} we use an adaptation of the Hadamard-WKB method developed by Anderson and Hu \cite{Anderson:2003} to efficiently calculate the coordinate series expansion of $V(x,x')$ to very high order for both Nariai and Schwarzschild spacetimes. In Sec.~\ref{sec:convergence} we use convergence tests to determine the radius of convergence of our series and show that, as expected, it lies within the convex normal neighborhood. We also estimate the local truncation error arising from truncating the series at a specific order. Using the method of Pad\'{e} approximants, we show in Sec.~\ref{sec:Pade} how the domain of validity of the coordinate series can be extended beyond its radius of convergence to give an accurate representation of $V(x,x')$ to within a small distance of the edge of the normal neighborhood.

\section{Hadamard-WKB Calculation of the Green Function}
\label{sec:WKB}

For the present quasilocal calculation, we need to consider the retarded Green function only for the points $x$ and $x'$ lying within a \emph{normal neighborhood}. This allows us to express the retarded Green function in the Hadamard parametrix \cite{Hadamard,Friedlander},
\begin{equation}
\label{eq:Hadamard}
G_{ret}\left( x,x' \right) = \theta_{-} \left( x,x' \right) \left\lbrace U \left( x,x' \right) \delta \left( \sigma \left( x,x' \right) \right) - V \left( x,x' \right) \theta \left( - \sigma \left( x,x' \right) \right) \right\rbrace ,
\end{equation}
where $\theta_{-} \left( x,x' \right)$ is analogous to the Heaviside step-function (i.e. $1$ when $x'$ is in the causal past of $x$, $0$ otherwise), $\delta \left( \sigma\left( x,x' \right) \right)$ is the standard Dirac delta function, $U \left( x,x' \right)$ and $V \left( x,x' \right)$ are symmetric bi-scalars having the benefit that they are regular for $x' \rightarrow x$, and $\sigma \left( x,x' \right)$ is the Synge \cite{Synge,Poisson:2003,DeWitt:1965} world function (i.e., half the square of the geodesic distance). The term involving $U(x,x')$ is only non-zero for null connected points whereas the quasi-local self-force calculation which motivates us requires the Green function within the light-cone only. We will therefore only concern ourselves here with the calculation of the function $V(x,x')$.

The fact that $x$ and $x'$ are close together suggests that an expansion of $V(x,x')$ in powers of the separation of the points,
\begin{equation}
\label{eq:CoordGreen}
V\left( x,x' \right) = \sum_{i,j,k=0}^{\infty} v_{ijk}(r) \left( t - t' \right)^{2i} \left( \cos \gamma - 1 \right)^j (r-r')^k,
\end{equation}
where $\gamma$ is the angular separation of the points, 
may give a good representation of the function within the quasi-local region. Note that, as a result of the spherical symmetry of the spacetimes we will be considering, the expansion coefficients, $v_{ijk}(r)$, are only a function of the radial coordinate, $r$. Anderson and Hu \cite{Anderson:2003} have developed a Hadamard-WKB method for calculating these coefficients. They applied their method to the Schwarzschild case and subsequently found the coefficients to $14^{th}$ order using the \emph{Mathematica} computer algebra system \cite{Anderson:Eftekharzadeh:Hu:2006}. In the present work, we adapt their method to allow for spacetimes of the Nariai form, (\ref{eq:Nariai-le}). In particular, we consider a class of spacetimes of the general form
\begin{equation}
\label{eq:diagonal-metric}
ds^2 = -f(r) dt^2 + f^{-1}(r) dr^2 + g(r) \left( d\theta^2 + \sin^2\theta d\phi^2 \right),
\end{equation}
where $f(r)$ and $g(r)$ are arbitrary functions of the radial coordinate, $r$, and previously Anderson and Hu had set $g(r) = r^2$, but allowed $g_{rr}$ and $g_{tt}$ to be independent functions of $r$. The form of the Nariai metric given in Eq.~(\ref{eq:Nariai-le}) falls into the class \eqref{eq:diagonal-metric}, with $f(r) = 1- r^2$ and $g(r) = 1$. For $f(r) = 1-\frac{2M}{r}$ and $g(r)=r^2$, this is the Schwarzschild metric of Eq.~(\ref{eq:Schw-le}).

The method presented in this section differs from that of Ref.~\cite{Anderson:2003} in the details of the WKB approach used, but otherwise remains very similar. Our alternative WKB approach, based on that of Refs.~\cite{Howard:1985,Winstanley:2007} proves extremely efficient when implemented in a computer algebra package.

Following the prescription of Ref.~\cite{Anderson:2003}, the Hadamard parametrix for the real part of the Euclidean Green function (corresponding to the Euclidean metric arising from the change of coordinate $\tau = i t$) is \footnote{Note that this definition of the Green function differs from that of Ref.~\cite{Anderson:2003} by a factor of $4\pi$ and the definition of $V(x,x')$ differs by a further factor of 2.}
\begin{equation}
\label{eq:HadamardEuclideanGF}
 \Re \left[ G_E (-i \tau,\vec{x};-i\tau',\vec{x}') \right]= \frac{1}{2\pi} \left( \frac{U(x,x')}{\sigma(x,x')} + V(x,x') \ln (|\sigma(x,x')|) + W(x,x') \right) ,
\end{equation}
where $U(x,x')$, $V(x,x')$ and $W(x,x')$ are real-valued symmetric bi-scalars.

Additionally, for the points $x$ and $x'$ separated farther apart in the time direction than in other directions,
\begin{equation}
 \sigma (x,x') = -\frac{1}{2} f(r) (t-t')^2 + O\left[(x-x')^3\right]
\end{equation}
so the logarithmic part of Eq.~(\ref{eq:HadamardEuclideanGF}) is given by
\begin{equation}
\label{eq:VlnTau}
 \frac{1}{\pi} V(x,x') \ln(\tau - \tau').
\end{equation}
Therefore, in order to find $V(x,x')$, it is sufficient to find the coefficient of the logarithmic part of the Euclidean Green function. We do so by considering the fact that the Euclidean Green function also has the exact expression for the spacetimes of the form given in Eq.~(\ref{eq:diagonal-metric}):
\begin{equation}
\label{eq:EuclideanGF}
  G_E (-i \tau,x;-i\tau',x') = \frac{1}{\pi} \int_{0}^{\infty} d\omega \cos \left[ \omega (\tau - \tau') \right] \sum_{l=0}^{\infty} (2l+1) P_l (\cos \gamma) C_{\omega l} p_{\omega l}(r_<) q_{\omega l}(r_>),
\end{equation}
where $p_{\omega l}$ and $q_{\omega l}$ are solutions (normalised by $C_{\omega l}$) to the homogeneous radial equation for the scalar wave
equation in the curved background (\ref{eq:diagonal-metric}) (and where $r_< $, $r_>$ are the smaller/larger of $r$ and $r'$, respectively), along with the fact that
\begin{align}
\label{eq:WKB-int-log-relation}
 \int_\lambda^{\infty} d\omega \cos \left[ \omega (\tau - \tau') \right] \frac{1}{\omega^{2n+1}} &= \frac{(-1)^{n+1}}{(2n)!}(\tau-\tau')^{2n}\log\left(\tau-\tau'\right) + \cdots \nonumber\\
	&= \frac{-1}{(2n)!}(t-t')^{2n}\log\left(\tau-\tau'\right)+\cdots ,
\end{align}
where $\lambda$ is a low frequency cut-off justified by the fact that we will only need the $\log(\tau-\tau')$ term from the integral. We can therefore find $V(x,x')$ as an expansion in powers of the time separation of the points by expressing the sum,
\begin{equation}
\label{eq:WKBsum}
\sum_{l=0}^{\infty} (2l+1) P_l (\cos \gamma) C_{\omega l} p_{\omega l}(r_<) q_{\omega l}(r_>),
\end{equation}
of Eq.~(\ref{eq:EuclideanGF}) as an expansion in inverse powers of $\omega$. This is achieved using a WKB-like method based on that of Refs.~\cite{Howard:1985,Winstanley:2007}. Given the form (\ref{eq:EuclideanGF}) for the Euclidean Green function, the radial functions $S(r)=p_{\omega l}(r)$ and $S(r)=q_{\omega l}(r)$ must both satisfy the homogeneous wave equation,
\begin{equation}
\label{eq:WKB-radial-eq}
 f \frac{d^2 S}{dr^2} + \frac{1}{g} \frac{d}{dr}(f g) \frac{dS}{dr} - \left[ \frac{\omega^2}{f} + \frac{l(l+1)}{g} + m_{\text{field}}^2 + \xi R\right] S = 0
\end{equation}
where $m_{\text{field}}$ is the scalar field mass and $\xi$ is the coupling to the scalar curvature, $R$. Next, given the Wronskian $W(r) = C_{\omega l}(p_{\omega l}q_{\omega l}' - q_{\omega l}p_{\omega l}')$, its derivative is
\begin{equation}
 W' = C_{\omega l}(p_{\omega l}q_{\omega l}'' - q_{\omega l}p_{\omega l}'') = - \frac{1}{fg} (fg)' W
\end{equation}
and the Wronskian condition is therefore
\begin{equation}
\label{eq:WKBWronskian}
 C_{\omega l}(p_{\omega l}q'_{\omega l} - q_{\omega l}p'_{\omega l}) = -\frac{1}{fg}.
\end{equation}
We now explicitly assume that $r>r'$ and define the function
\begin{equation}
 B(r,r') = C_{\omega l} p_{\omega l}(r') q_{\omega l}(r).
\end{equation}
Since the sum of Eq.~(\ref{eq:WKBsum}) (and hence $B(r,r')$) is only needed as an expansion in powers of $(r-r')$, we expand $B(r,r')$ about $r'=r$ and (using Eq.~(\ref{eq:WKB-radial-eq}) to replace second order and higher derivatives of $B(r,r')$ with expressions involving $B(r,r')$ and $\partial_{r'}B(r,r')$) find that
\begin{equation}
 B(r,r') = \beta(r) + \alpha(r) (r'-r) + \left\{ \left[\frac{2 (\eta+\chi^2)}{(fg)^2}\right] \beta(r) - \left[\ln(fg) \right]' \alpha(r)  \right\} \frac{(r'-r)^2}{2} + \cdots\qquad(\textrm{for\ }r > r')
\end{equation}
where
\begin{align}
\beta(r) \equiv& \left[ B(r,r') \right]_{r' \to r^-} = C_{\omega l} p_{\omega l}(r) q_{\omega l}(r),
& \alpha(r) \equiv \left[\partial_{r'}B(r,r')\right]_{r' \to r^-}=&  C_{\omega l} p'_{\omega l}(r) q_{\omega l}(r)
\end{align}
and 
\begin{align}
\label{eq:eta_defn}
 \eta(r) &\equiv -\frac{1}{4}f g+\left(m_{\text{field}}^2+\xi  R\right) f g^2\\
 \label{eq:chi_defn}
 \chi^2(r) &\equiv \omega^2 g^2 +  f g \left(l+\frac{1}{2}\right)^2 .
\end{align}
It will therefore suffice to calculate $\beta(r)$ and $\alpha(r)$. Furthermore, using Eq.~\eqref{eq:WKBWronskian} we can relate $\alpha(r)$ to the derivative of $\beta(r)$,
\begin{equation}
 \alpha (r) = \frac{\beta'(r)}{2} + \frac{1}{2f(r)g(r)},
\end{equation}
so it will, in fact, suffice to find $\beta(r)$ and its derivative, $\beta'(r)$.

Using Eqs.~(\ref{eq:WKB-radial-eq}) and (\ref{eq:WKBWronskian}), it is immediate to see that $\beta(r)$ must satisfy the nonlinear differential equation
\begin{equation} \label{eq:beta-ode}
 f g \frac{d}{dr}\left(f g \frac{d\sqrt{\beta}}{dr}\right)-\left(\eta +\chi ^2\right) \sqrt{\beta}+ \frac{1}{4 \beta^{3/2}}=0.
\end{equation}

The short distance behaviour of the Green function is determined by the high-$\omega$ and/or high-$l$ behaviour of the
integrand of Eq.~\eqref{eq:EuclideanGF}, so we seek to express $\beta(r)$ as an expansion in inverse powers of $\chi$.  To keep track of this 
expansion we may replace  $\chi$ in Eq.~(\ref{eq:beta-ode}) by $\chi/\epsilon$ where $\epsilon$ is a formal expansion 
parameter which we eventually set to 1.
Then, to balance at leading order we require
\begin{align}
(\chi/\epsilon)^2 \sqrt{\beta} \sim \frac{1}{4 \beta^{3/2}} \quad \implies \beta \sim \frac{\epsilon}{2 \chi} .
\end{align}
We now write 
\begin{align} \label{eq:beta}
\beta(r)&= \epsilon \beta_0(r) + \epsilon^2\beta_1(r) + \ldots
\end{align}
where $\beta_0(r) \equiv 1/(2 \chi(r))$, insert this form for $\beta(r)$ in Eq.~(\ref{eq:beta-ode}), and solve formally order by order in $\epsilon$ to find a recursion relation for the $\beta_n(r)$. 
On doing so and using  Eq.~(\ref{eq:chi_defn}) to eliminate $(l+\frac12)^2$ in favour of $\omega^2$ and $\chi^2$ we find
that we can write
\begin{align} \label{eq:beta-n}
\beta_n(r) = \sum\limits_{m=0}^{2n} \frac{A_{n,m}(r) \omega^{2m} }{\chi^{2n+2m+1}}
\end{align}
so, for example,
\begin{align}   
\beta_1(r) &=\frac{A_{1,0}(r)}{\chi^3}+\frac{A_{1,1}(r) \omega^2}{\chi^5}+\frac{A_{1,2}(r)\omega^4}{\chi^7}\\
\beta_2(r) &=\frac{A_{2,0}(r)}{\chi^5}+\frac{A_{2,1}(r) \omega^2}{\chi^7}+\frac{A_{2,2}(r)\omega^4}{\chi^9}+\frac{A_{2,3}(r) \omega^6}{\chi^{11}}+\frac{A_{2,4}(r)\omega^8}{\chi^{13}}. 
\end{align}
The recursion relations for $\beta_n(r)$ may then be re-expressed to allow us to recursively solve for the $A_{n,m}(r)$. Such a recursive calculation is ideally suited to implementation in a computer algebra system (CAS). Even on a computer, this recursive calculation becomes very long as $n$ becomes large and, in fact, dominates the time required to calculate the series expansion of $V(x,x')$ as a whole. For this reason, we have made available an example implementation in \emph{Mathematica}, including precalculated results for several spacetimes of interest \cite{Hadamard-WKB-Code}. This code calculates analytic results
for $A_{n,m}(r)$ on a moderate Linux workstation up to order $n\sim 25$, corresponding to a separation $|x-x'|^{50}$, in Schwarzschild and Nariai space-times in the order of 1~hour of CPU time.

We also note at this point that knowledge of the $A_{n,m}(r)$ (and their $r$ derivative, which is straightforward to calculate) are all that is required to find the series expansion of $\beta'(r)$ and hence $\alpha(r)$. This can be seen by differentiating Eqs.~\eqref{eq:beta} and \eqref{eq:beta-n} with respect to $r$ to get
\begin{align}
 \beta'(r) &= \epsilon \beta'_0 + \epsilon^2 \beta'_1 + \ldots,
\end{align}
with
\begin{align}\label{eq:betap-n}
 \beta'_n(r) &= \sum_{m=0}^{2n} \left[ \frac{A'_{n,m}(r) \omega^{2m} }{\chi^{2n+2m+1}} - (n+m+\frac12) \frac{2 \chi \chi' A_{n,m}(r)}{\chi^{2n+2m+3}} \right]\nonumber\\
	&=  \sum_{m=0}^{2n+1} \left\{ A'_{n,m}(r)  - (n+m+\frac12) A_{n,m}(r)\frac{(fg)'}{fg} - (n+m-\frac12)A_{n,m-1}(r)\left[ (g^2)'-\frac{(fg)'}{fg}g^2\right]\right\} \frac{\omega^{2m} }{\chi^{2n+2m+1}}
\end{align}
where we have used Eq.~\eqref{eq:chi_defn} to write $2 \chi \chi' = \chi^2 (fg)'/(fg) + \omega^2 ((g^2)' -g^2 (fg)'/(fg))$, and we use the convention $A_{n,-1}=A_{n,2m+1} = 0$.

With the $A_{n,m}(r)$ and their first derivatives calculated, we are faced with the sum over $l$ in Eq.~(\ref{eq:WKBsum}), where Eqs.~\eqref{eq:beta-n} and \eqref{eq:betap-n} yield sums of the form
\begin{equation}
\label{eq:WKB-simplified-sum}
 \sum_{l=0}^{\infty} 2 (l+{\textstyle\frac{1}{2}}) P_l (\cos \gamma) \frac{D_{n,m}(r) \omega^{2m}}{\chi^{2n+2m+1}}.
\end{equation}
with
\begin{equation}
 D_{n,m} = \begin{cases} A_{n,m}(r)\\
A'_{n,m}(r)  - (n+m+\frac12) A_{n,m}(r)\frac{(fg)'}{fg} - (n+m-\frac12)A_{n,m-1}(r)\left[ (g^2)'-\frac{(fg)'}{fg}g^2\right]
\end{cases}
\end{equation}

Since we are considering the points $x$ and $x'$ to be close together, we can treat $\gamma$ as a small quantity and expand the Legendre polynomial in a Taylor series about $\gamma=0$, or, more conveniently, in powers of $(\cos \gamma -1)$ about $(\cos \gamma -1)=0$. It is straightforward to express each term in this series as a polynomial in even powers of $(l+\frac{1}{2})$:
\begin{align}
P_l(\cos \gamma) = {}_2F_1\left(-l,l+1;1;(1-\cos\gamma)/2\right) =\sum_{p=0}^l \frac{\bigl((l+\frac12)^2-(1-\frac12)^2\bigr)\cdots
\bigl((l+\frac12)^2-(p-\frac12)^2\bigr)}{2^p (p!)^2} (\cos\gamma-1)^p
\end{align}
 The calculation of the sum in Eq.~(\ref{eq:EuclideanGF}) therefore reduces to the calculation of sums of the form
\begin{equation}
\label{eq:WKBsum2}
2 D_{n,m}(r) \sum_{l=0}^{\infty}  \frac{(l+{\textstyle\frac{1}{2}})^{2p+1} \omega^{2m}}{\chi^{2n+2m+1}}.
\end{equation}
For fixed $\omega$ and large $l$ the summand behaves as $l^{2(p-n-m)}$, and so only converges if $p<n+m$.
If $p \geq n+m$, we first split the summand as
\begin{align}
\frac{(l+{\textstyle\frac{1}{2}})^{2p+1} \omega^{2m}}{\chi^{2m+2n+1}} =&\frac{(l+{\textstyle\frac{1}{2}})^{2(p-m-n)}\omega^{2m}}{(fg)^{m+n+1/2}}\left(1+\frac{\omega^2 g/f}{   \left(l+\frac{1}{2}\right)^2} \right)^{-m-n-1/2}\\
=&\frac{(l+{\textstyle\frac{1}{2}})^{2(p-m-n)}\omega^{2m}}{(fg)^{m+n+1/2}}\left\{ \sum_{k=0}^{p-m-n}
\frac{(-1)^k(m+n+1/2)_k}{k!} \left(\frac{\omega^2 g/f}{   \left(l+\frac{1}{2}\right)^2} \right)^{k}
+\right.\nonumber \\
&+ \left. \left[ \left(1+\frac{\omega^2 g/f}{   \left(l+\frac{1}{2}\right)^2} \right)^{-m-n-1/2}-\sum_{k=0}^{p-m-n}
\frac{(-1)^k (m+n+1/2)_k}{k!} \left(\frac{\omega^2 g/f}{   \left(l+\frac{1}{2}\right)^2} \right)^{k}
\right]\right\}
\end{align}
where $(\alpha)_k=\Gamma(\alpha+k)/\Gamma(\alpha)$ is the Pochhammer symbol and the sum correspond to the first $(p-n-m)$ terms in the expansion of $ (1+x)^{-n-m-1/2}$ about $x\equiv(\omega^2 g/f)/(l+\frac12)^2=0$.
The terms outside the square brackets correspond to positive powers of $\omega$ and so contribute to the light cone singularity, not the tail term $V(x,x')$ with which we are concerned in this paper. By contrast, the term in square brackets behaves
as $(l+\frac12)^{-2}$ and so converges as $l\to\infty$ and will contribute to the tail term. We denote this term, with its
prefactor as
\begin{align}
\label{eq:reg_integrand}
\left[\frac{(l+{\textstyle\frac{1}{2}})^{2p+1} \omega^{2m}}{\chi^{2m+2n+1}}\right]_\mathrm{reg} 
&=
\frac{\omega^{2(p-n)}}{f^{p+1/2}g^{2(n+m)-p+1/2}} x^{m+n-p}
\left(  \left(1+ x \right)^{-m-n-1/2} -\sum_{k=0}^{p-m-n}
\frac{(-1)^k (m+n+1/2)_k}{k!} x^{k} \right)
\end{align}
and adopt the understanding that the sum vanishes if $p<m+n$.
 
To proceed further, we use the Sommerfeld-Watson formula \cite{Watson:1918},
\begin{equation}
\sum_{l=0}^{\infty} F(l+{\textstyle\frac12}) = \Re  \left[ \frac{1}{i} \int_\gamma dz \> F(z) \tan (\pi z) \right] = \int_{0}^{\infty} F\left(\lambda\right) d\lambda - \Re \left( i \int_{0}^{\infty} \frac{2}{1+e^{2\pi \lambda}} F\left(i \lambda \right) d\lambda \right)
\end{equation}
which is valid provided we can rotate the contour of integration for $F(z) \tan(\pi z)$ from just above the real axis to the positive imaginary axis.
Defining $z \equiv (f/g)^{1/2} \lambda/\omega = 1/\sqrt{x}$, where $\lambda\equiv (l+1/2)$, the sum \eqref{eq:WKBsum2} can then be written as the contour integral,
\begin{equation}
\label{eq:WKBintegrals}
 \frac{\omega^{2(p-n)+1}}{f^{p+1}g^{2(m+n)-p}}\left[
\int_{0}^{\infty} \mathrm{d}z\>\left[ \frac{ z^{2p+1} }{(1 +z^2)^{m+n+1/2}}\right]_\mathrm{reg}  
+ (-1)^p \Re \left( \int_{0}^{\infty} \frac{2\mathrm{d}z}{1+e^{2\pi z \omega \sqrt{g/f}}} \frac{ z^{2p+1}}{(1 - z^2)^{m+n+1/2}} \right)\right]
\ .
\end{equation}
Note that there is no need to include the regularization terms in the second integral as their contribution is manifestly imaginary and so will not contribute to the final answer.

For $p<m+n$ the first integral in Eq.~(\ref{eq:WKBintegrals}) may be performed immediately as
\begin{equation} 
\int_{0}^{\infty} \mathrm{d}z\> \frac{ z^{2p+1} }{(1 +z^2)^{m+n+1/2}} =  \frac{p!}{2 (m+n-p-1/2)_{p+1}} \ .
\end{equation}
For $p\geq m+n$, we use the regularised integrand arising from Eq.~(\ref{eq:reg_integrand}) and temporarily introduce an ultraviolet cutoff $1/\epsilon^2$ to get
\begin{equation}  
 \int_{\epsilon}^{\infty} \mathrm{d}x \> x^{m+n-p-3/2}\left[(1+x)^{-m-n-1/2} -  \sum_{k=0}^{p-m-n}
\frac{(-1)^k (m+n+1/2)_k}{k!} x^{k} \right]\ .
\end{equation}
Integrating by parts $p-m-n+1$ times, the regularisation subtraction terms ensure that boundary term go to zero in the limit $\epsilon\to 0$ and we are left with
 \begin{equation}  
(-1)^{p-m-n+1} \frac{(m+n+1/2)_{p-m-n+1}}{(1/2)_{p-m-n+1}}\int_{\epsilon}^{\infty} \mathrm{d}x \> x^{-1/2} (1+x)^{-p-3/2}
=
 \frac{(-1)^{p-m-n+1} \pi p!}{\Gamma(m+n+1/2)\Gamma(p-m-n+3/2)}\>,
\end{equation}
where the last equality reflects that after these integrations by parts the
remaining integral is finite in the limit $\epsilon\to 0$.
 
\begin{figure}[ht]
 \begin{center}
  \includegraphics[width=6cm]{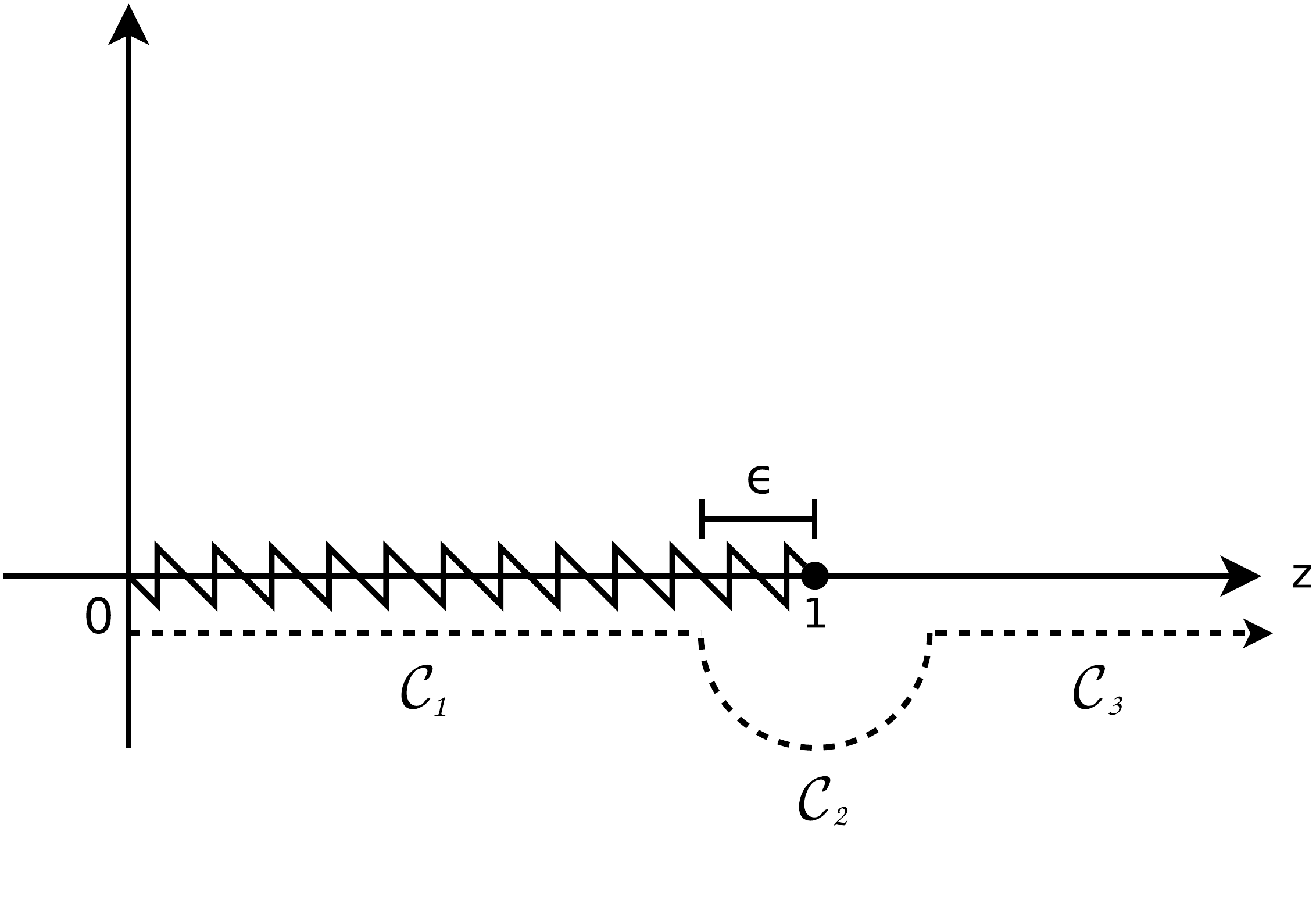}
 \end{center}
 \caption{The second integral in Eq.~(\ref{eq:WKBintegrals}) has a pole at $z=1$, so we split it into three parts: (1) An integral from $0$ to $1-\epsilon$, (2) An arc of radius $\epsilon$ about $z=1$ (3) An integral from $z=1+\epsilon$ to $\infty$.}
 \label{fig:wkbcontour}
\end{figure}

The second integral in Eq.~(\ref{eq:WKBintegrals}) is understood as a contour integral as illustrated in Fig.~\ref{fig:wkbcontour}. The integrand is understood to be defined on the complex plane cut from $z=-1$ to $1$
and additionally possesses singularities at $z=-1$ and $z=1$. To handle these in a fashion consistent with the Watson-Sommerfeld prescription we consider the contour in 3 parts: $\mathcal{C}_1$ running just below the cut from $0$ to $1-\epsilon$,
 $\mathcal{C}_2$ a semicircle of radius $\epsilon$ about $z=1$, and $\mathcal{C}_3$ running along the real axis from $z=1+\epsilon$ to $\infty$.

The integrand along $\mathcal{C}_3$ is manifestly imaginary and so gives zero contribution.
Writing the integrand as
\begin{equation}
\frac{ G(z)}{(1 - z)^{m+n+1/2}}
\end{equation}
where
\begin{equation}
\label{eq:Gdef}
 G(z) = \frac{z^{2p+1}}{(1+e^{2\pi z \omega \sqrt{g/f}})(1+z)^{m+n+1/2}}
\end{equation}
it is straightforward to see that
\begin{align}
\Re \int_{\mathcal{C}_2} \frac{ G(z)}{(1 - z)^{m+n+1/2}} \mathrm{d}z = - \sum\limits_{k=0}^\infty \frac{(-1)^{k}}{k!} G^{(k)}(1) 
\frac{\epsilon^{k-m-n+1/2}}{k-m-n+\frac{1}{2}}\ ,
\end{align}
while integrating by parts $m+n$ times
\begin{align}
\int_{\mathcal{C}_1} \frac{ G(z)}{(1 - z)^{m+n+1/2}} \mathrm{d}z =& \sum\limits_{k=0}^{m+n-1} \frac{(-1)^{k}}{k!} G^{(k)}(1) 
\frac{\epsilon^{k-m-n+1/2}-1}{k-m-n+\frac{1}{2}}\nonumber \\
& + \int\limits_0^{1-\epsilon} \frac{ \mathrm{d} z}{(1-z)^{m+n+1/2}}
  \left[  G(z) - \sum\limits_{k=0}^{m+n-1} \frac{(-1)^{k}}{k!} G^{(k)}(1) (1-z)^k \right]\ .
\end{align}
Adding the contributions from these components, it is clear that the $\epsilon \to 0$ divergences cancel and we are left with
\begin{align}
\label{eq:Gresult}
- \sum\limits_{k=0}^{m+n-1} \frac{(-1)^{k}}{k!} \frac{G^{(k)}(1)}{k-m-n+\frac{1}{2}} + \int\limits_0^{1} \frac{ \mathrm{d} z}{(1-z)^{m+n+1/2}}
  \left[  G(z) - \sum\limits_{k=0}^{m+n-1} \frac{(-1)^{k}}{k!} G^{(k)}(1) (1-z)^k \right]\ ,
\end{align}
where the subtraction terms in the integrand ensure the integral here is well-defined.

While we can take this analysis further~\cite{Ottewill:Winstanley:Young:2009}, 
for our current purpose we note that we only need the expansion of Eq.~(\ref{eq:Gresult}) in terms of an inverse \textit{powers} of $\omega$
as  $\omega\to \infty$.
From Eq.~(\ref{eq:Gdef}), it is immediate that the terms in the sum in (\ref{eq:Gresult}) are exponentially small
and so may be ignored for our purposes here.  Indeed we may simultaneously increase the upper limit
on the two sums in Eq.~(\ref{eq:Gresult}) without changing the result.  Increasing it by 1 (or more), the
integand increases from 0, peaks and then decreases to 0 at 1 with the peak approaching 0 as $\omega\to\infty$.
Standard techniques from statistical mechanics then dictate that the $\omega\to\infty$ asymptotic form of
the integral follows from the expanding the integrand, aside from the `Planck factor',
about $z=0$ and extending the upper limit to $\infty$. Again, in doing so the contribution from the summation within the integrand give exponentially small contribution
so that the powers of $\omega$ are determined simply by
\begin{align}
\int\limits_0^{\infty} \frac{ \mathrm{d} z}{(1+e^{2\pi z \omega \sqrt{g/f}})}
   \mathop{\mathrm{Series}}\limits_{z=0} \left[\frac{z^{2p+1}}{(1-z^2)^{m+n+1/2}}\right]
\end{align}
The coefficients in the series are known analytically, so to expand in inverse powers of $\omega$ we only need to compute integrals of the form
\begin{equation}
\int_0^{\infty} \frac{z^{2N-1}}{1+e^{2\pi   z \omega \sqrt{g/f}}} \, dz
\end{equation}
which have the exact solutions \cite{GradRyz}
\begin{equation}
\left(1-2^{1-2N}\right)\frac{f^N}{g^N \omega^{2N}}\frac{|B_{2N}|}{4N},
\end{equation}
where $B_N$ is the $N$-th Bernoulli number. This expression allows us to calculate the integrals very quickly.

Applying this method for summation over $l$, Eq.~(\ref{eq:EuclideanGF}) takes the form required by Eq.~(\ref{eq:WKB-int-log-relation}) so we now have the logarithmic part of the Euclidean Green function and therefore $V(x,x')$ as the required power series in $(t-t')$, $(\cos \gamma - 1)$ and $(r-r')$. Due to the length of the expressions involved, we have made available online \cite{Hadamard-WKB-Code} a \emph{Mathematica} code implementing this algorithm, along with precalculated results for several spacetimes of interest including Schwarzschild, Nariai and Reissner-Nordstr\"om.

\section{Convergence of the Series}
\label{sec:convergence}
We have expressed $V(x,x')$ as a power series in the separation of the points. This series will, in general, not be convergent for all point separations -- the maximum point separation for which the series remains convergent will be given by its \emph{radius of convergence}. In this section, we explore the radius of convergence of the series in the Nariai and Schwarzschild spacetimes and use this as an estimate on the region of validity of our series.

For simplicity, we will consider points separated only in the time direction so we will have a power series in $(t-t')$,
\begin{equation}
\label{eq:CoordGreenT}
V\left( x,x' \right) = \sum_{n=0}^{\infty} v_{n}(r) \left( t - t' \right)^{2n}.
\end{equation}
where $v_{n}(r)$ is a real function of the radial coordinate, $r$ only. We will also consider cases where the points are separated by a fixed amount in the spatial directions, or where the separation in other directions can be re-expressed in terms of a time separation, resulting in a similar power series in $(t-t')$, but with the coefficients, $v_{n}(r)$, being different. This will give us sufficient insight without requiring overly complicated convergence tests.

In the next section, we review some tests that will prove useful. In Secs.~\ref{sec:Nariai-tests} and \ref{sec:Schw-tests} we present the results of applying those tests in Nariai and Schwarzschild spacetimes, respectively.
\subsection{Tests for Estimating the Radius of Convergence} \label{sec:Tests}
\subsubsection{Convergence Tests}
For the power series (\ref{eq:CoordGreenT}), there are two convergence tests which will be useful for estimating the radius of convergence. The first of these, the \emph{ratio test}, gives an estimate of the radius of convergence, $\Delta t_{RC}$,
\begin{equation}
 \Delta t_{RC} = \lim_{n\rightarrow\infty}\sqrt{\left|\frac{v_{n}}{v_{n+1}}\right|}.
\end{equation}
Although strictly speaking, the large $n$ limit must be taken, in practice we have calculated enough terms in the series to get a good estimate of the limit by simply looking at the last two terms. The ratio test falls into difficulties, however, when one of the terms in the series is zero. Unfortunately, this occurs frequently for many cases of interest. It is possible to avoid this issue somewhat by considering non-adjacent terms in the series, i.e. by comparing terms of order $n$ and $n+m$, 
\begin{equation}
\label{eq:ratio-test}
 \Delta t_{RC} = \lim_{n\rightarrow\infty}\left|\frac{v_{n}}{v_{n+m}}\right|^{\frac{1}{2m}}.
\end{equation}
Using the ratio test in this way gives better estimates of the radius of convergence, although the results are still somewhat lacking.

To get around this difficulty, we also use a second test, the \emph{root test},
\begin{equation}
  \Delta t_{RC} = \limsup_{n\rightarrow\infty} \left|\frac{1}{v_n}\right|^{\frac{1}{2n}},
\end{equation}
which is well suited to power series. This gives us another estimate of the radius of convergence. Again, in practice the last calculated term in the series will give a good estimate of the limit.

It may appear that only the (better behaved) root test is necessary for estimating the radius of convergence of the series. However, extra insight can be gained from including both tests. This is because for the power series \eqref{eq:CoordGreenT}, the ratio test typically gives values \emph{increasing} in $n$ while the root test gives values \emph{decreasing} in $n$, effectively giving lower and upper bounds on the radius of convergence.

\subsubsection{Normal Neighborhood} \label{subsubsec:NN}
The Hadamard parametrix for the retarded Green function, (\ref{eq:Hadamard}), is only guaranteed to be valid provided $x$ and $x'$ are within a \emph{normal neighborhood} (see footnote \ref{def:causal domain}). This arises from the fact that the Hadamard parametrix involves Synge's world function, $\sigma(x,x')$, which is only defined for the points $x$ and $x'$ separated by a \emph{unique geodesic}. It is therefore plausible that the radius of convergence of our series could exactly correspond to the normal neighborhood size, $t_{\text{NN}}$. This turns out to not be the case, although it is still helpful to give consideration to $t_{\text{NN}}$ as it should place an upper bound on the radius of convergence of the series.

The normal neighborhood size will be given by the minimum time separation of the spacetime points such that they are connected by two geodesics. For typical cases of interest for self-force calculations there will be a particle following a time-like geodesic, so $t_{\text{NN}}$ will be given by the minimum time taken by a null geodesic intersecting the particle's world-line twice. In typical black hole spacetimes, this geodesic will orbit the black hole once before re-intersecting the particle's world-line.

Another case of interest is that of the points $x$ and $x'$ at constant spatial points, separated by a constant angle $\Delta\phi$. Initially, (i.e. when $t=t'$), the points will be separated only by spacelike geodesics. After sufficient time has passed for a null geodesic to travel between the points (going through and angle $\Delta\phi$, they will be connected first by a null geodesic and subsequently by a sequence of unique timelike geodesics. Since the geodesics are unique, $t_{\text{NN}}$ will not be given by this \emph{first} null geodesic time. Rather, $t_\text{NN}$ will be given by the time taken by the \emph{second} null geodesic (passing through an angle $2\pi - \Delta\phi$). This subtle, but important, distinction will be clearly evident when we study specific cases in the next sections.

\subsubsection{Relative Truncation Error}
Knowledge of the radius of convergence alone does not give information about the accuracy of the series representation of $V(x,x')$. The series is necessarily truncated after a finite number of terms, introducing a \emph{truncation error}. As was done previously in Refs.~\cite{Ottewill:Wardell:2008,Anderson:Wiseman:2005} the local fractional truncation error can be estimated by the ratio between the highest order term in the expansion ( $O \left( \Delta \tau ^n \right)$, say) and the sum of all the terms up to that order,
\begin{equation} \label{eq:trunc-error}
\epsilon \equiv \frac{f^a_{\rm QL}\left[ n \right]}{\sum_{i=0}^{n} f^a_{\rm QL}\left[ i \right]}.
\end{equation}
Refs.~\cite{Ottewill:Wardell:2008,Anderson:Wiseman:2005} considered only the first two terms in the series when producing these estimates. Since we now have a vastly larger number of terms available, it is worthwhile considering these again to determine the accuracy of the high order series.

\subsection{Nariai Spacetime} \label{sec:Nariai-tests}
\subsubsection{Normal Neighborhood} \label{subsec:Nariai-NN}

Allowing only the time separation of the points to change, there are two cases of interest for the Nariai spacetime:
\begin{enumerate}
 \item The static particle which has a normal neighborhood determined by the minimum coordinate time taken by a null geodesic circling the origin ($\rc=0$) before returning. This is the time taken by a null geodesic, starting at $\rc=\rc_1$ and returning to $\rc'=\rc_1$, while passing through an angle $\Delta\phi=2\pi$. \label{enum:static-nariai}
 \item Points at fixed radius, $\rc_1$, separated by an angle, say $\pi/2$. In this case, there is a null geodesic which goes through an angle $\Delta\phi=\pi/2$ when traveling between the points. However, there will not yet be any other geodesic connecting them, so this will not give $t_{\text{NN}}$. Instead, it is the the next null geodesic, which goes through $\Delta\phi=3\pi/2$, that gives the normal neighborhood boundary. \label{enum:static-nariai-angle}
\end{enumerate}

In both cases, the coordinate time taken by the null geodesic to travel between the points is given by \cite{Casals:Dolan:Ottewill:Wardell:2009}
\begin{equation}
 t_{NN} = 2 \tanh^{-1}(\rc_1) + \ln \left( \frac{1-\rc_1 \sech^2(\Delta\phi) + \tanh(\Delta\phi) \sqrt{1 - \rc_1^2 \sech^2(\Delta\phi)}}{ 1+\rc_1 \sech^2(\Delta\phi) - \tanh(\Delta\phi) \sqrt{1 - \rc_1^2 \sech^2(\Delta\phi) }}  \right)  \label{eq:Nariai-geodesic-time}.
\end{equation}

\subsubsection{Results}
For the Nariai spacetime, the ratio test suffers from difficulties arising from zeros of the terms in the series. This is because the $i^{th}$ term of the series (of order $(t-t')^{2i}$) has $2i$ roots in $\rho$. In other words, the higher order the terms considered, the more likely one of the coefficients is to be near a zero, and not give a useful estimate of the radius of convergence. We have therefore compared non-adjacent terms to avoid this issue as much as possible, by choosing $m=n/2$ in Eq.~\eqref{eq:ratio-test}. Fortunately, the root test is much less affected by such issues and can be used without any adjustments.

\begin{figure}
  \begin{center}
  \includegraphics[width=6cm]{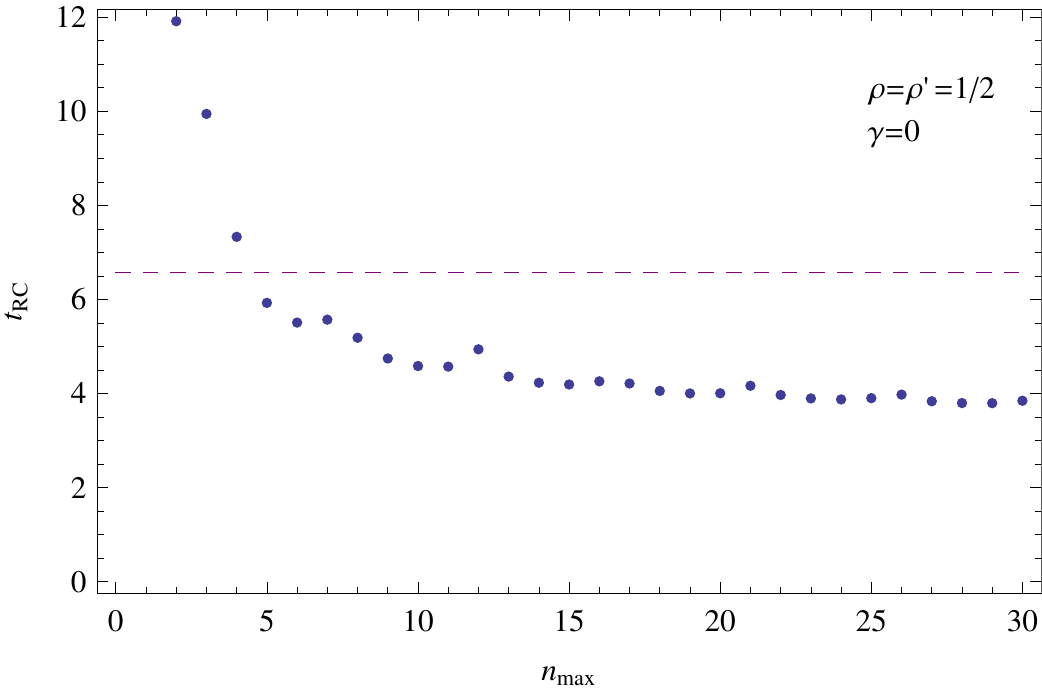}
  \includegraphics[width=6cm]{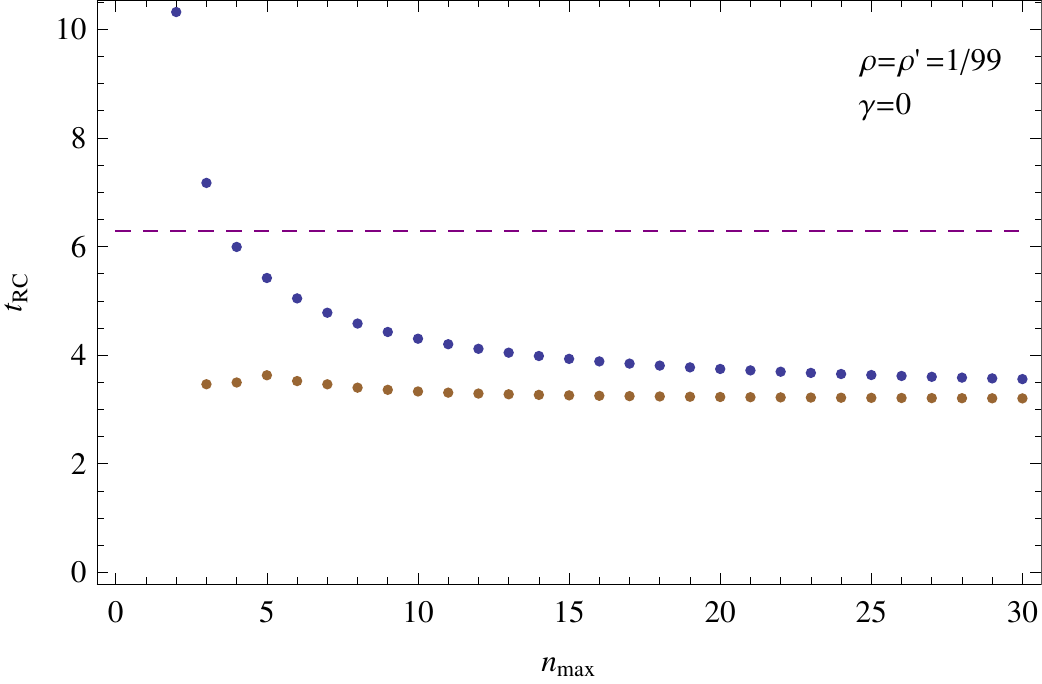}
  \includegraphics[width=6cm]{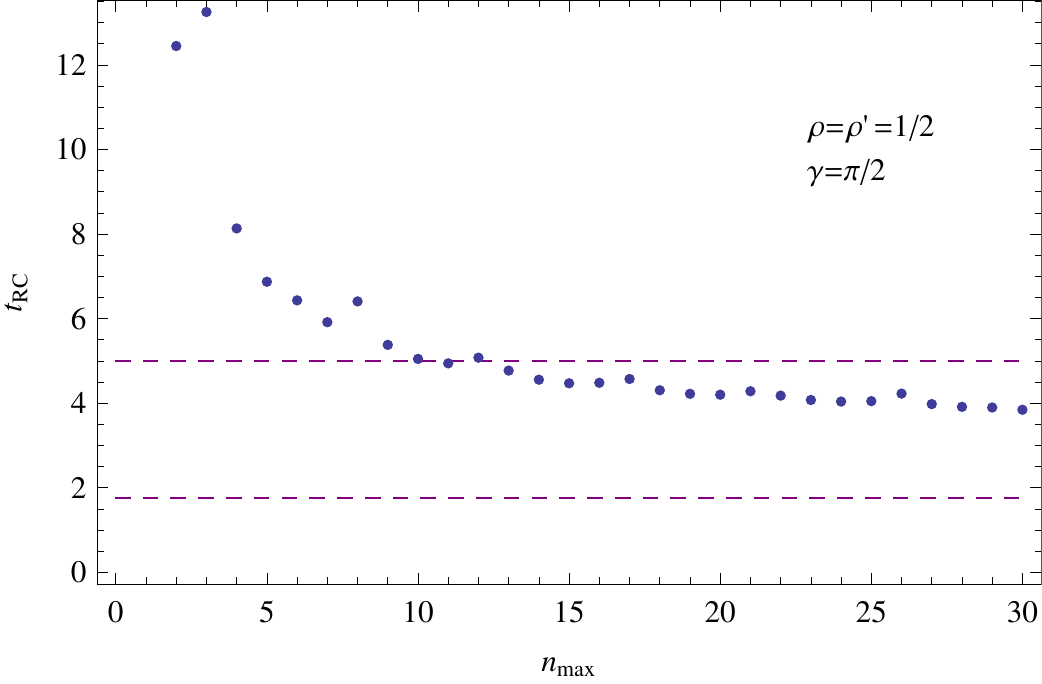}
  \includegraphics[width=6cm]{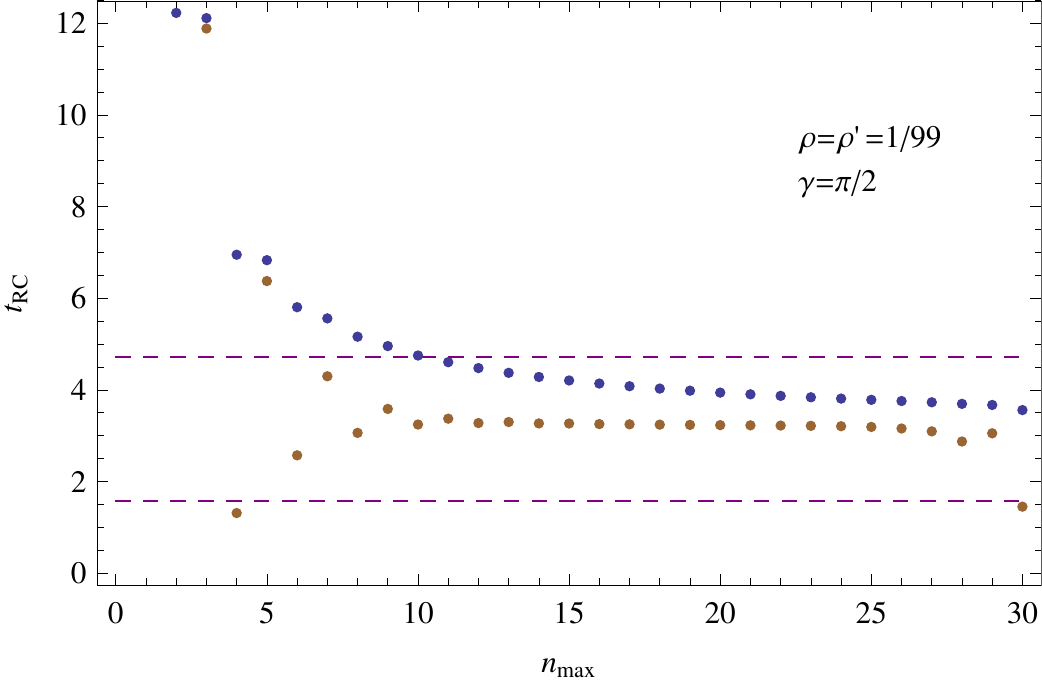}
\end{center}
 \caption{\emph{Radius of convergence as a function of the number of terms in the series for the Nariai spacetime with curvature coupling $\xi=1/6$.} The limit as $n\rightarrow\infty$ will give the actual radius of convergence, but it appears that just using the terms up to $n=30$ is giving a good estimate of this limit. The radius of convergence is estimated by the root test (blue dots) and ratio test (brown dots) and is compared against the normal neighborhood size (purple dashed line) calculated from considerations on null geodesics (see Sec.~\ref{subsubsec:NN} and Sec.~\ref{subsec:Nariai-NN}). Note that plots for the ratio test were omitted in cases where it did not give meaningful results.}
 \label{fig:nariai-roc2}
\end{figure}

In Fig.~\ref{fig:nariai-roc2}, we fix the radial position of the points at $\rho=\rho'=1/2$ and $\rho=\rho'=1/99$ and plot the results of applying the root test (blue dots) and ratio test (brown dots) for cases (\ref{enum:static-nariai}) Static particle (left) and (\ref{enum:static-nariai-angle}) Points at fixed spatial points separated by an angle $\gamma=\pi/2$ (right) as a function of the maximum order, $n_\text{max}$, of the terms of the series considered, $v_\text{nmax} \left(t-t^\prime \right)^{2n_\text{max}}$. It seems in both cases that the plot is limiting towards a constant value for the radius of convergence. In the case (\ref{enum:static-nariai}), we see that this gives a radius of convergence that is considerably smaller than the normal neighborhood size (purple dashed line). For the case (\ref{enum:static-nariai-angle}), we again see that the root test is limiting to a value for the radius of convergence. However, as discussed in Sec.~\ref{subsubsec:NN}, it is not the first null geodesic (lower dashed purple line), but the second null geodesic (upper dashed purple line) that determines the normal neighborhood and places an upper bound on the radius of convergence of the series.

\begin{figure}
  \begin{center}
  \includegraphics[width=6cm]{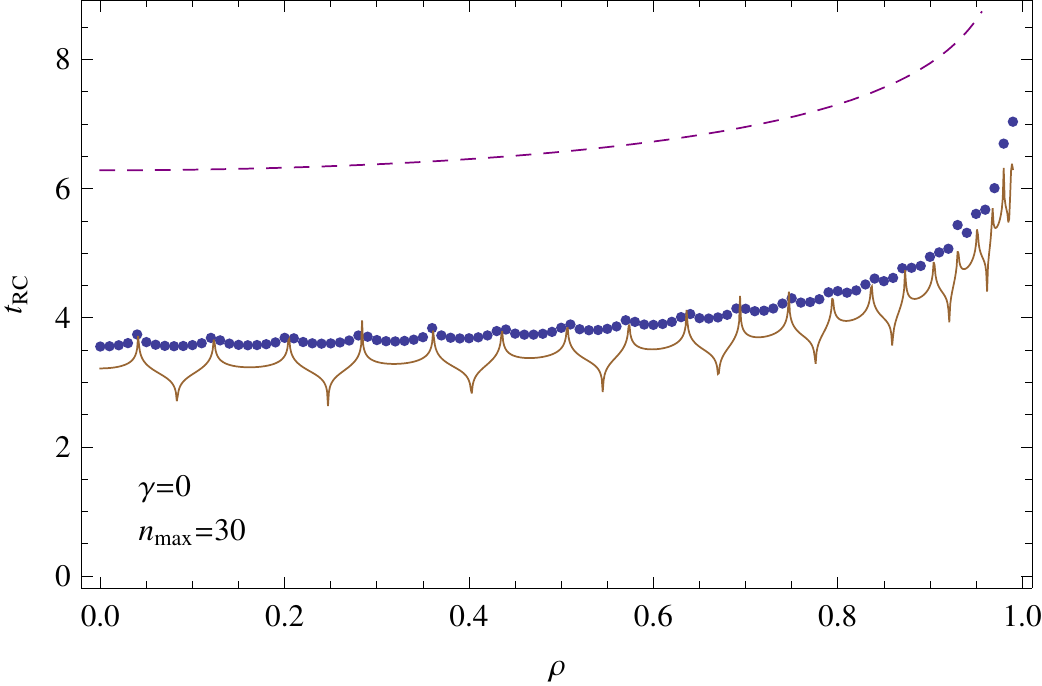}
  \includegraphics[width=6cm]{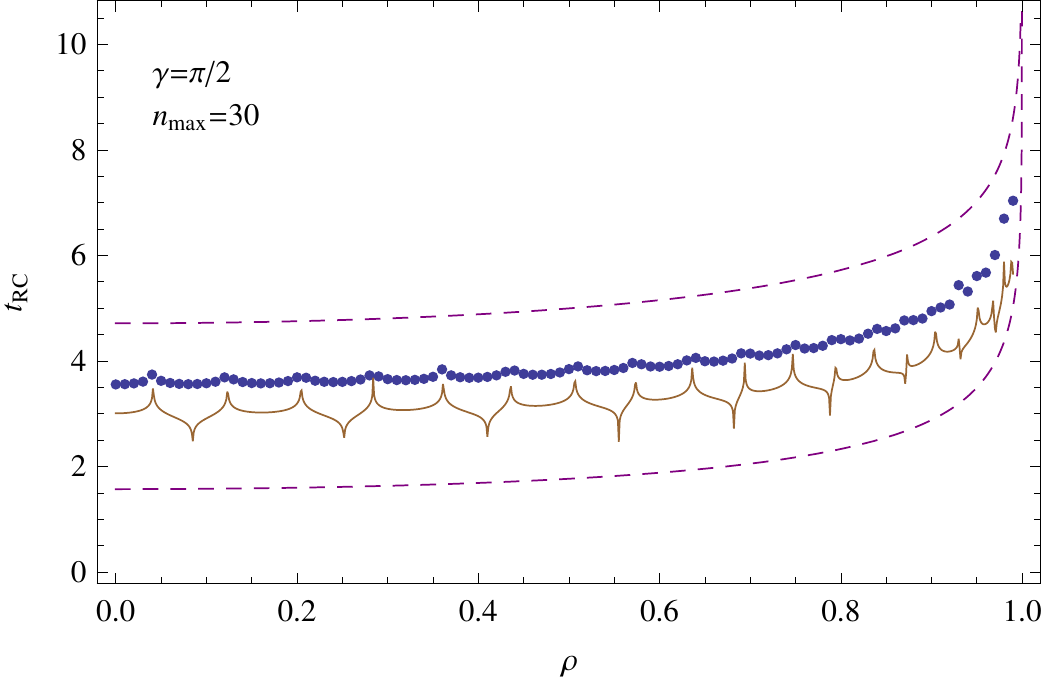}
\end{center}
 \caption{\emph{Estimates of the domain of validity (i.e. the radius of convergence) of the series expansion of $V(x,x')$ as a function of radial position in Nariai spacetime.} Root test on $O\left[(t-t')^{60}\right]$ series given as blue dots, ratio test given by brown line (see footnote \ref{fn:blips}) and normal neighborhood estimate from null geodesics given by dashed purple lines. The left plot is for the case (\ref{enum:static-nariai}), the static particle, right plot for the case (\ref{enum:static-nariai-angle}), points separated by an angle of $\pi/2$. In both cases, the series is clearly divergent before the normal neighborhood boundary. This boundary is sometimes given by the second, rather than first null geodesic as can be seen in the plot on the right. In particular, this is the case for the points separated by an angle $\gamma=\pi/2$ since they are initially separated by only spacelike geodesics (see Sec.~\ref{subsec:Nariai-NN}).}
 \label{fig:nariai-roc}
\end{figure}

In Fig.~\ref{fig:nariai-roc} we use the root test (blue dots) and ratio test (brown line)\footnote{\label{fn:blips}Note that the `blips' in the ratio test are an artifact of the zeroes of the series coefficients used and are not to be taken to have any physical meaning. In fact, the `blips' occur at different times when considering the series at different orders, so they should be ignored altogether. The ratio test plots should therefore only be fully trusted \emph{away} from the `blips'. Near the blips, it is clear that one could interpolate an approximate value, however we have not done so here as the plot is to be taken only as an \emph{indication} of the radius of convergence.} to investigate how the the radius of convergence of the series varies as a function of the radial position of the points, $\rc_1$. Again, we look at both cases
(\ref{enum:static-nariai}) (left) and (\ref{enum:static-nariai-angle}) (right). As a reference, we compare to the normal neighborhood size (purple dashed line). We find that, regardless of the radial position of the points, the radius of convergence of the series is well within the normal neighborhood, by an almost constant amount. As before, in the case (\ref{enum:static-nariai-angle}) of the points separated by an angle, we find that it is the second, not the first null geodesic that gives the normal neighborhood size.

\begin{figure}
  \begin{center}
  \includegraphics[width=6cm]{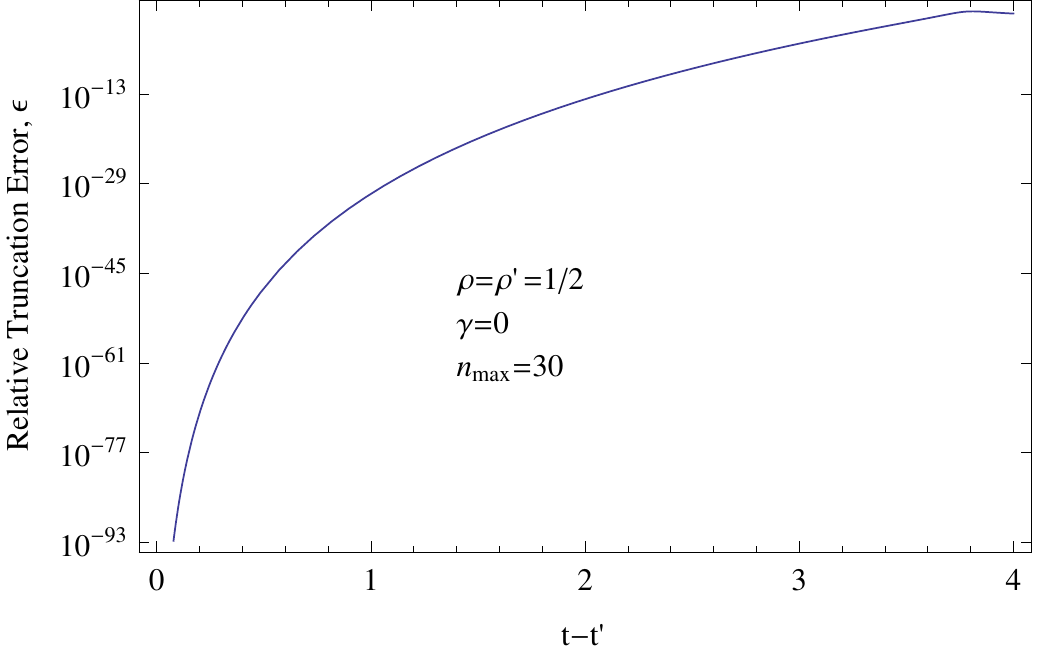}
  \includegraphics[width=6cm]{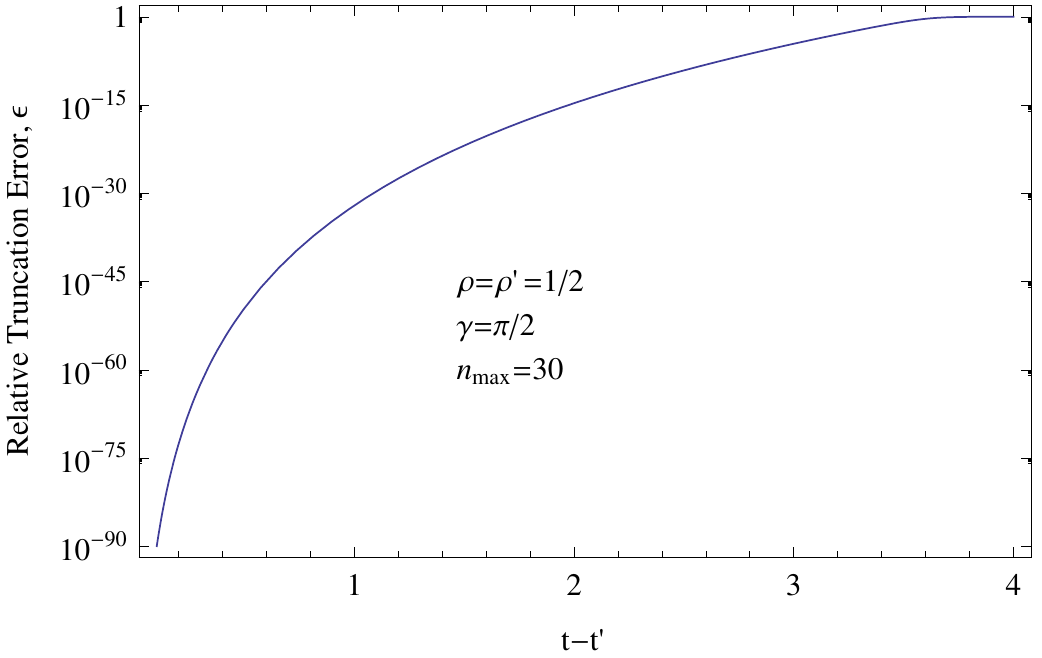}
\end{center}
 \caption{\emph{Relative truncation error in Nariai spacetime} arising from truncating the series expansion for $V(x,x')$ at order $|x-x'|^{60}$ (i.e. $n_\text{max}=30$) for cases (\ref{enum:static-nariai}) the static particle (left) and (\ref{enum:static-nariai-angle}) points separated by an angle $\pi/2$ (right). In both cases, the radial points are fixed at $\rho=\rho'=1/2$. The series is extremely accurate until we get close to the radius of convergence (see Fig.\ref{fig:nariai-roc2}).}
 \label{fig:nariai-trunc}
\end{figure}

With knowledge of radius of convergence of the series established, it is also important to estimate the accuracy of the series within that radius. To that end, we plot in Fig.~\ref{fig:nariai-trunc} the relative truncation error, (\ref{eq:trunc-error}), as a function of the time separation of the points at a fixed radius, $\rho=\rho'=1/2$. We find that the $60^{th}$ order series is extremely accurate to within a short distance of the radius of convergence of the series.

\subsection{Schwarzschild Spacetime} \label{sec:Schw-tests}
\subsubsection{Normal Neighborhood}
For a fixed spatial point at radius $r_1$ in the Schwarzschild spacetime, we would like to find the null geodesic that intersects it twice in the shortest time. This geodesic will orbit the black hole once before returning to $r_1$. Clearly, the coordinate time $t_{\text{NN}}$ for this orbit can only depend on $r_1$. The periapsis radius, $r_p$, will be reached half way through the orbit. For the radially inward half of this motion, the geodesic equations can be rearranged to give
\begin{align}
 \pi = \int_0^\pi d\phi &= - \int_{r_1}^{r_p} \frac{dr}{r^2 \sqrt{\frac{1}{r_p^2}\left(1-\frac{2M}{r_p}\right)-\frac{1}{r^2}\left(1-\frac{2M}{r}\right)}}\\
 \frac{t_{NN}}{2} = \int_0^{t_{NN}/2} dt &= - \int_{r_1}^{r_p} \frac{dr}{\left(1-\frac{2M}{r}\right)\sqrt{1-\frac{r_p^2}{r^2}\left(1-\frac{2M}{r}\right)\left(1-\frac{2M}{r_p}\right)^{-1}}}
\end{align}
For a given point $r_1$, we numerically solve the first of these to find the periapsis radius, $r_p$, then solve the second to give the normal neighborhood size, $t_{\text{NN}}$.

\subsubsection{Results}
The ratio test proves more stable for Schwarzschild than it was for Nariai. The series coefficients still have a large number of roots in $r$, but most are within $r=6M$ and therefore don't have an effect for the physically interesting radii, $r\ge6M$.

\begin{figure}
  \begin{center}
  \includegraphics[width=6cm]{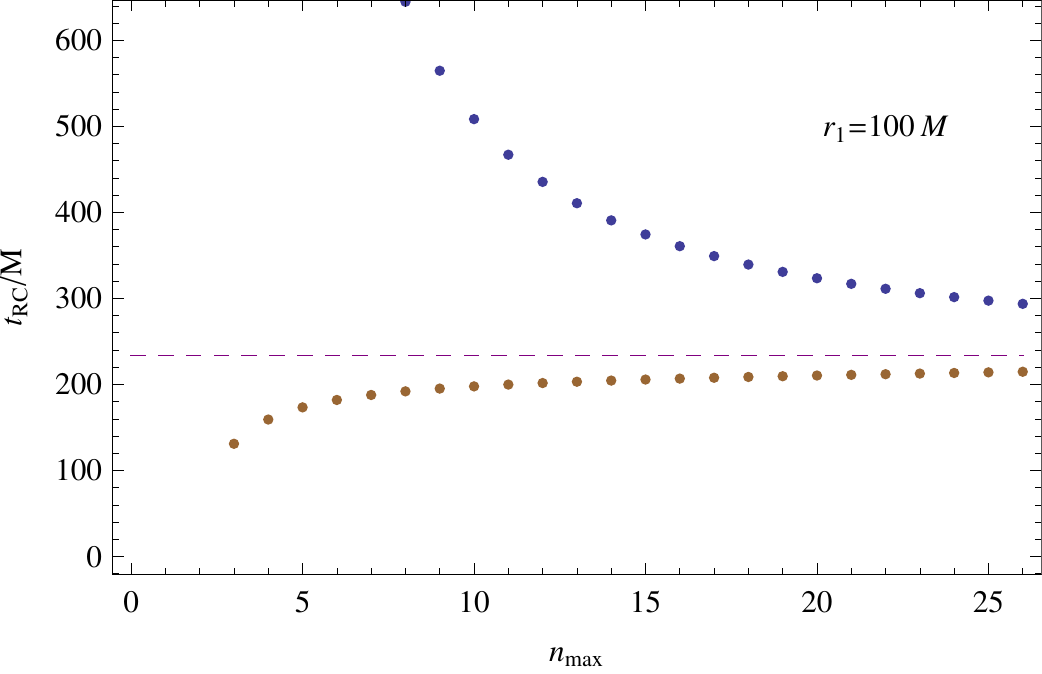}
  \includegraphics[width=6cm]{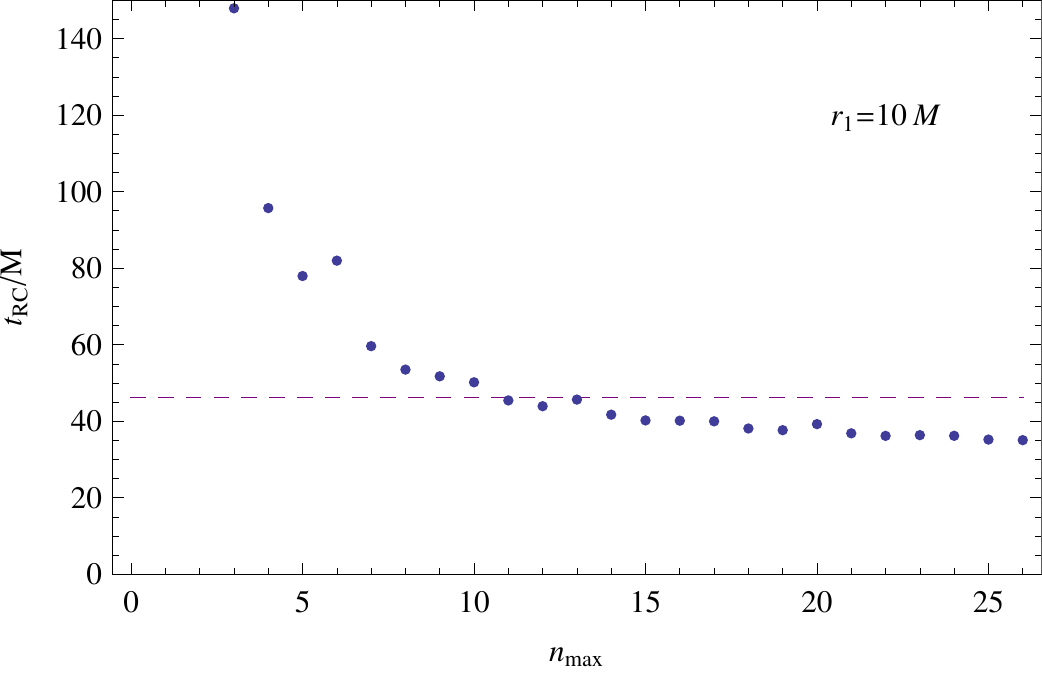}
 \end{center}
 \caption{\emph{Radius of convergence as a function of the number of terms considered.} Root test (blue dots), ratio test (brown line -- see footnote \ref{fn:blips}), first null geodesic (purple dashed line). Left: Static particle at $r_1=100M$ Right: Static particle at $r_1=10M$. The radius of convergence is given by the limit as the number of terms $n_{\text{max}} \to \infty$. It is apparent that the plots are asymptoting to a value near, but slightly lower than the normal neighborhood size. Note that curves for the ratio test were omitted in cases where it did not give meaningful results.}
 \label{fig:schw-static-roc-terms}
\end{figure}

Fig.~\ref{fig:schw-static-roc-terms} shows that the radius of convergence $ \Delta t_{RC}$ given by the root test is a decreasing function of the order of the term used, while that given by the ratio test is increasing. This effectively gives an upper and lower bound on the radius of convergence of the series. There is some `noise' in the ratio test plot at lower radii (where that test is failing to give meaningful results), but we simply ignore this and omit the ratio test in this case. For the root test, the terms up to order $(t-t')^{52}$ was used, while for the ratio test, adjacent terms in the series up order $(t-t')^{52}$ were compared.

\begin{figure}
  \begin{center}
  \includegraphics[width=6cm]{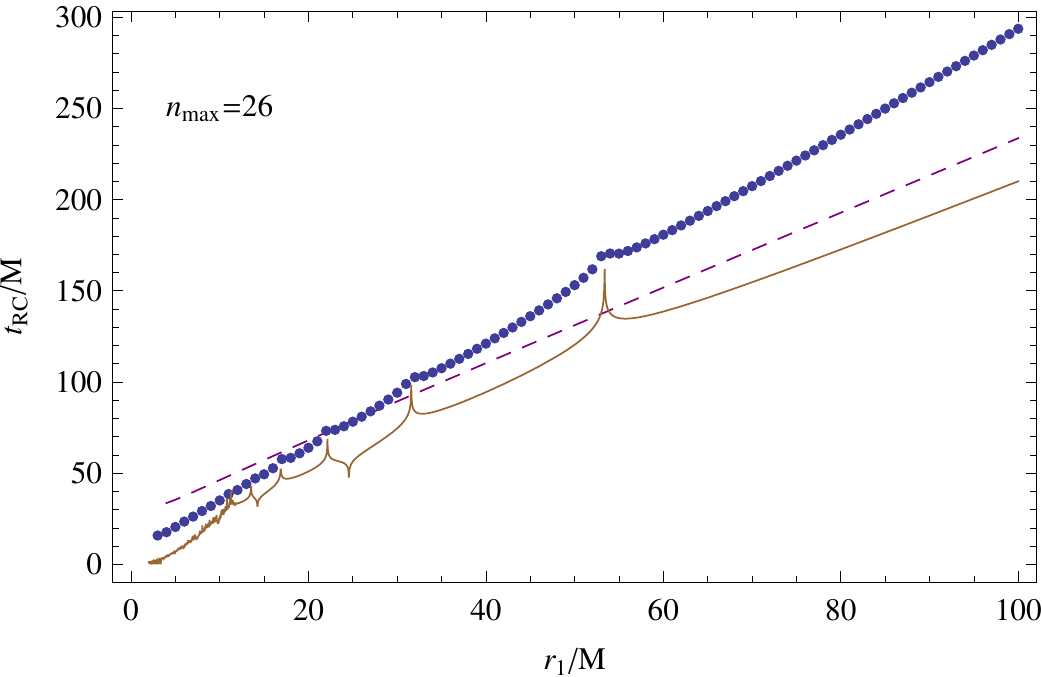}
 \end{center}
 \caption{\emph{Radius of convergence as a function of radial position for a static point in Schwarzschild.} Root test (blue dots), ratio test (brown line -- see footnote \ref{fn:blips}), first null geodesic (purple dashed line).}
 \label{fig:schw-static-roc-radius}
\end{figure}

In Fig. \ref{fig:schw-static-roc-radius} we apply the root (blue dots) and ratio (brown line -- see footnote \ref{fn:blips}) tests for the case of a static particle at a range of radii in Schwarzschild spacetime. Using the root test as an upper bound and the ratio test as a lower bound, it is clear that the radius of convergence is near, but likely slightly lower than the normal neighborhood size (purple dashed line). In this case, for the root test, the term of order $(t-t')^{52}$ was used, while for the ratio test, terms of order $(t-t')^{52}$ and $(t-t')^{26}$ were compared.

\begin{figure}
  \begin{center}
  \includegraphics[width=6cm]{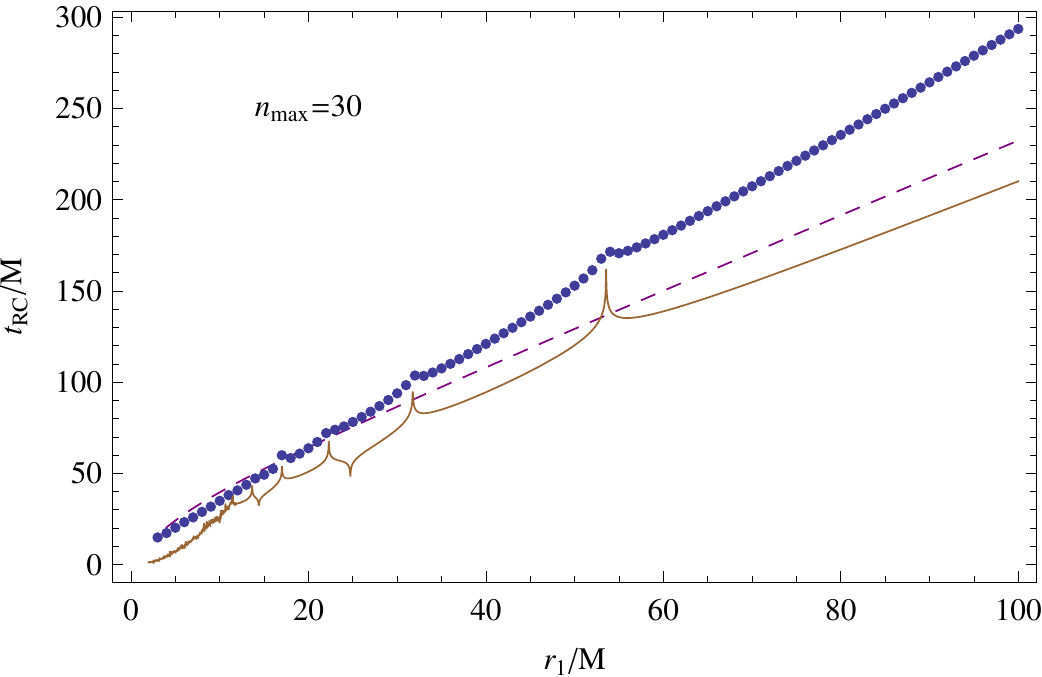}
  \includegraphics[width=6cm]{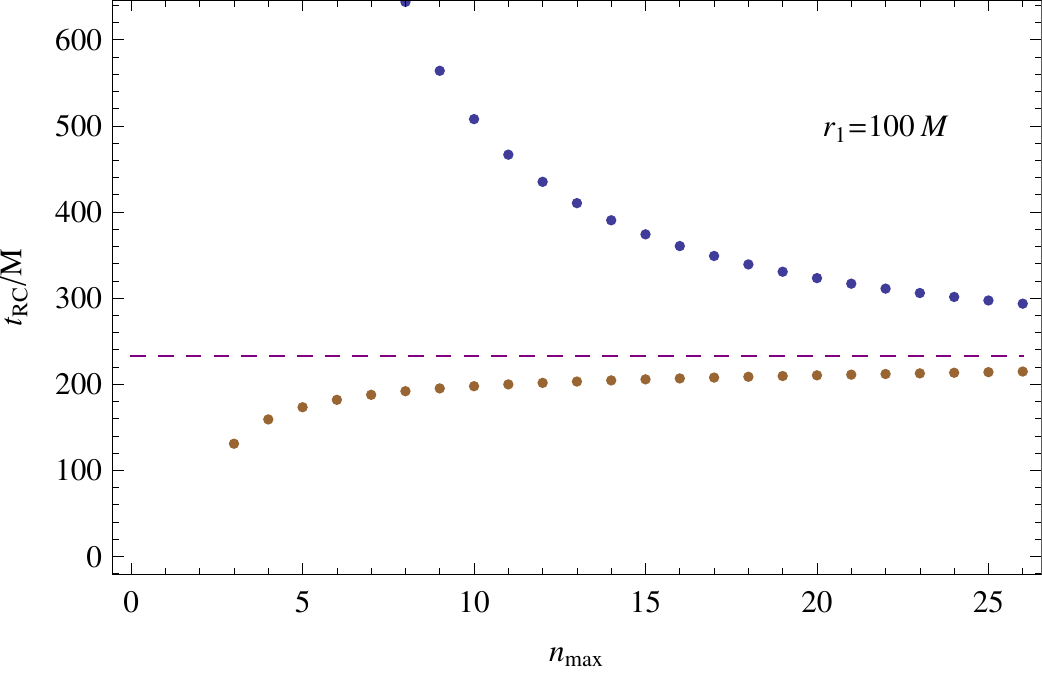}
  \includegraphics[width=6cm]{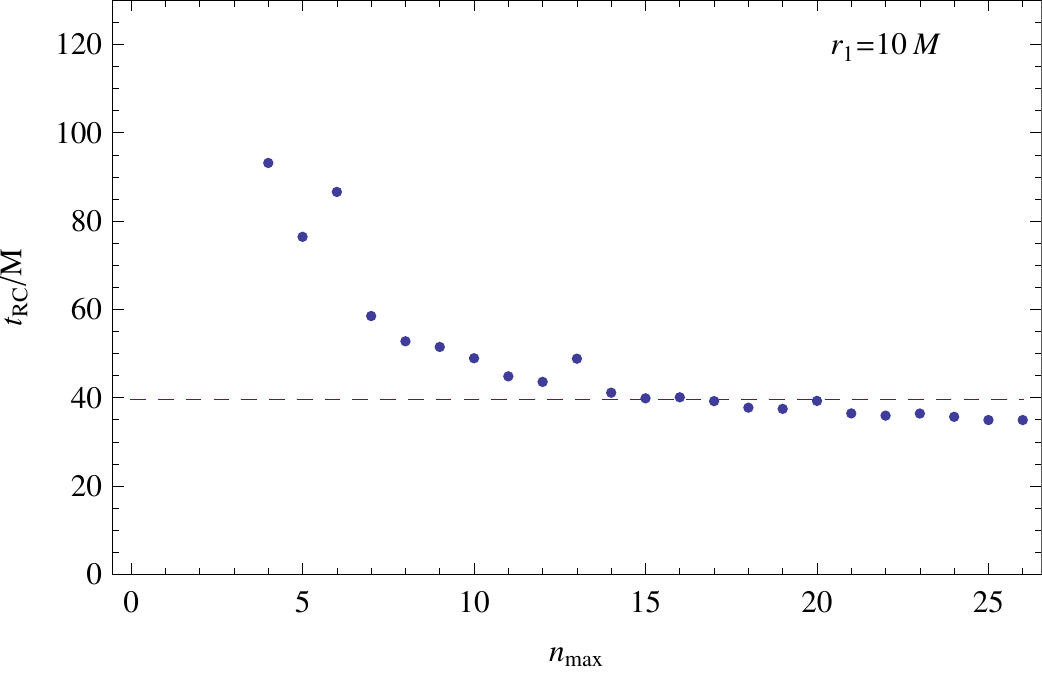}
 \end{center}
 \caption{\emph{Radius of convergence as a function of radial position for a points separated along a circular geodesic in Schwarzschild.} Root test (blue dots), ratio test (brown line -- see footnote \ref{fn:blips}), first null geodesic (purple dashed line). Note that curves for the ratio test were omitted in cases where it did not give meaningful results.}
 \label{fig:schw-circ-roc}
\end{figure}

In Fig.~\ref{fig:schw-circ-roc}, we repeat for the case of the points separated on a circular timelike geodesic. The results are very similar to the static particle case and show the same features.

\section{Extending the Domain of Series Using Pad\'{e} Approximants} \label{sec:Pade}
In the previous section it was shown that the circle of convergence of the series expansion of $V(x,x')$ is smaller than the size of the normal neighborhood. This is not totally unexpected. We would expect the normal neighborhood size to place an upper limit on the radius of convergence, but we can not necessarily expect the radius of convergence to be exactly the normal neighborhood size. However, since the Hadamard parametrix for the Green function is valid everywhere within the normal neighborhood, it is reasonable to hope that it would be possible to find an alternative series representation for $V(x,x')$ which is valid in the region outside the circle of convergence of the original series, while remaining within the normal neighborhood.

The radius of convergence found in the previous section locates the distance (in the complex plane) to the closest singularity of $V(x,x')$. However, that singularity could lie anywhere on the (complex) circle of convergence and will not necessarily be on the real line. In fact, given that the Green function is clearly not singular at the radius of convergence on the real line, it is clear that the singularity of $V(x,x')$
does not lie on the real line.

There are several techniques which can be employed to extend a series beyond its radius of convergence. Provided the circle of convergence does not constitute a \emph{natural boundary} of the function, the method of \emph{analytic continuation} can be used to find another series representation for $V(x,x')$ valid outside the circle of convergence of the original series \cite{M&F,Whittaker:Watson}. This could then be applied iteratively to find series representations covering the entire range of interest of $V(x,x')$. Although analytic continuation should be capable of extending the series expansion of $V(x,x')$, there is an alternative method, the method of \emph{Pad\'{e} approximants} which yields impressive results with little effort.

The method of Pad\'{e} approximants \cite{Bender:Orszag,NumericalRecipes} is frequently used to extend the series representation of a function beyond the radius of convergence of the series. It has been employed in the context of General Relativity data analysis with considerable success \cite{Damour:Iyer:Sathyaprakash,Porter:Sathyaprakash}. It is based on the idea of expressing the original series as a rational function (i.e. a ratio of two polynomials $V(x,x')=R(x,x')/S(x,x')$) and is closely related to the continued fraction representation of a function \cite{Bender:Orszag}. This captures the functional form of the singularities of the function on the circle of convergence of the original series. The Pad\'{e} approximant, $P_M^N (t-t')$ is defined as
\begin{equation}
 P_M^N (t-t') \equiv \frac{\sum_{n=0}^N A_n (t-t')^n}{\sum_{n=0}^M B_n (t-t')^n}
\end{equation}
where $B_0 = 1$ and the other $(M+N+1)$ terms are found by comparing to the first $(M+N+1)$ terms of the original power series. The choice of $M$ and $N$ is arbitrary provided $M+N\le n_{\text{max}}$ where $n_{\text{max}}$ is the highest order term that has been computed for the original series. There are, however, choices for $M$ and $N$ which give the best results. In particular, the diagonal, $P^N_N$, and sub-diagonal, $P^N_{N+1}$, Pad\'e approximants yield optimal results.

\subsection{Nariai} \label{subsec:padeNariai}
\begin{figure}
  \begin{center}
  \includegraphics[width=6cm]{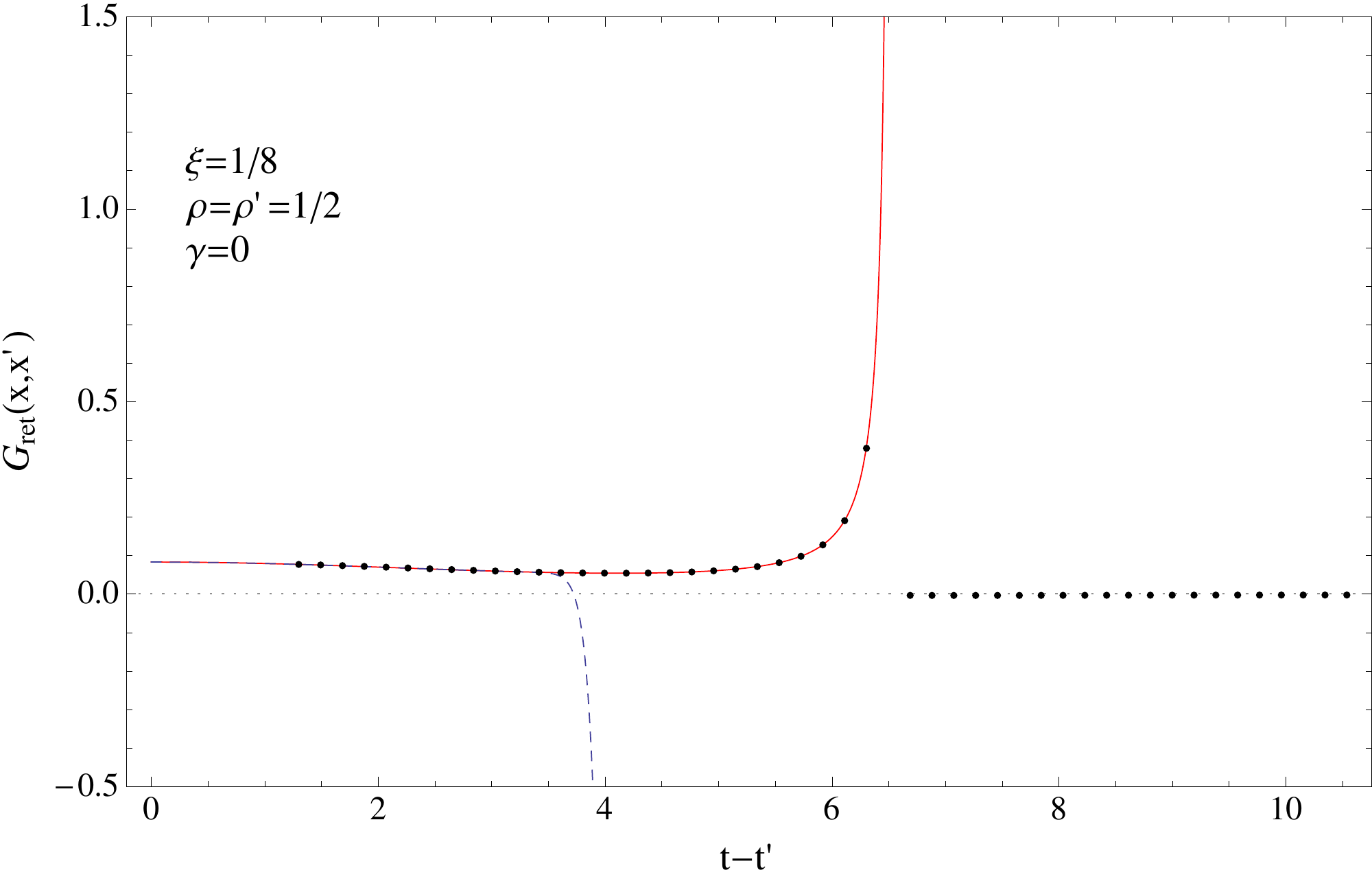}
  \includegraphics[width=6cm]{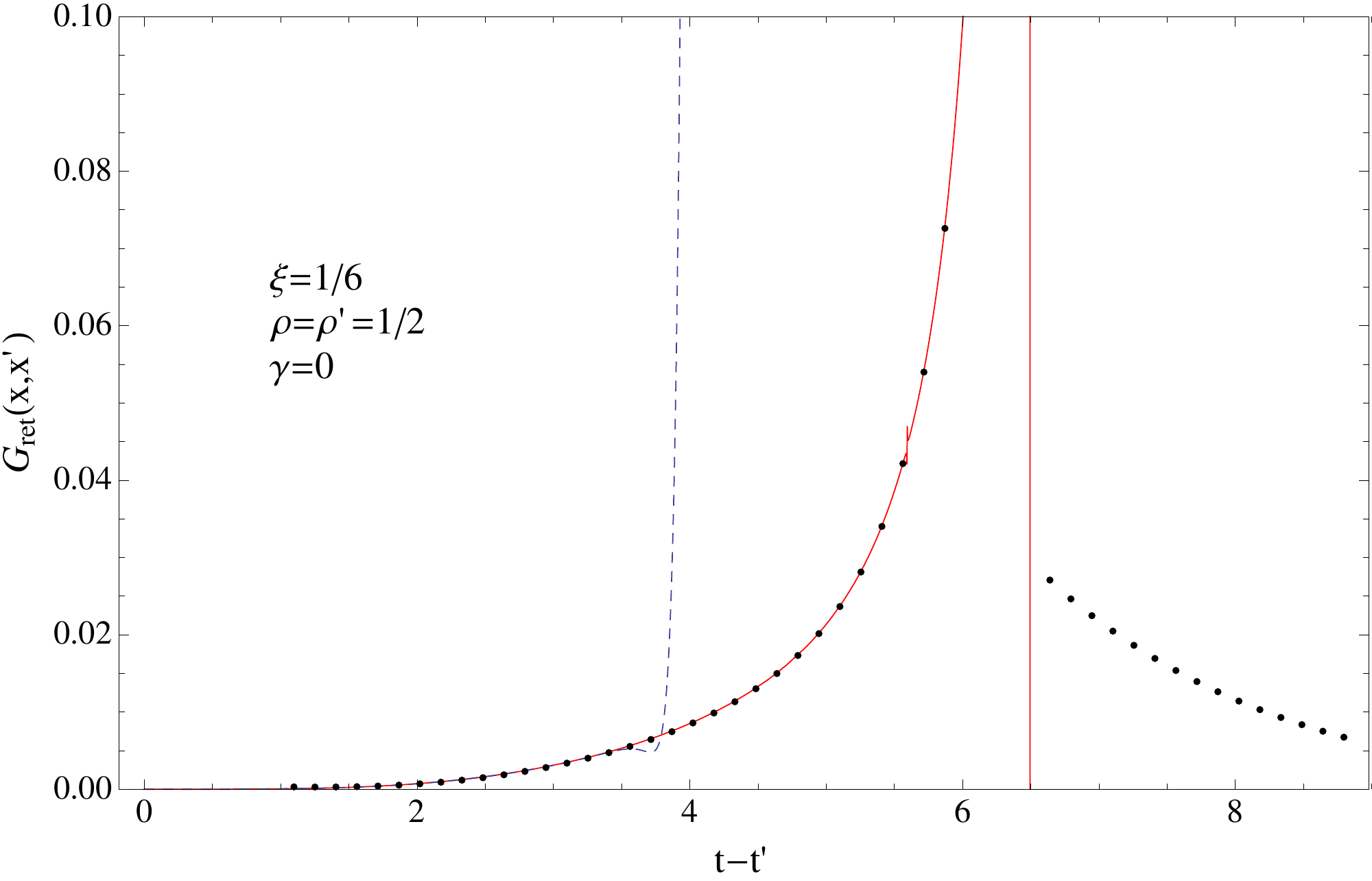}
  \includegraphics[width=6cm]{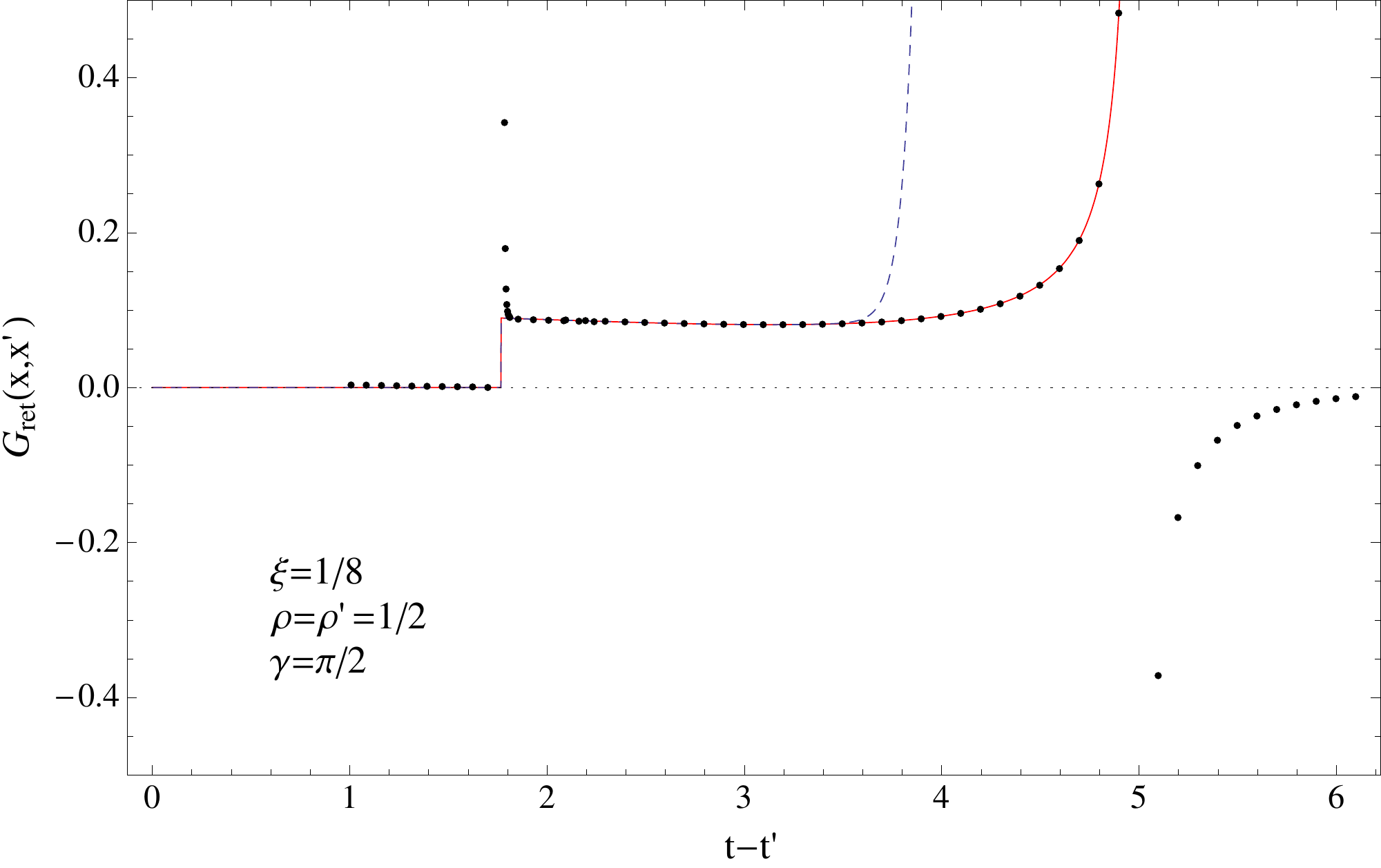}
  \includegraphics[width=6cm]{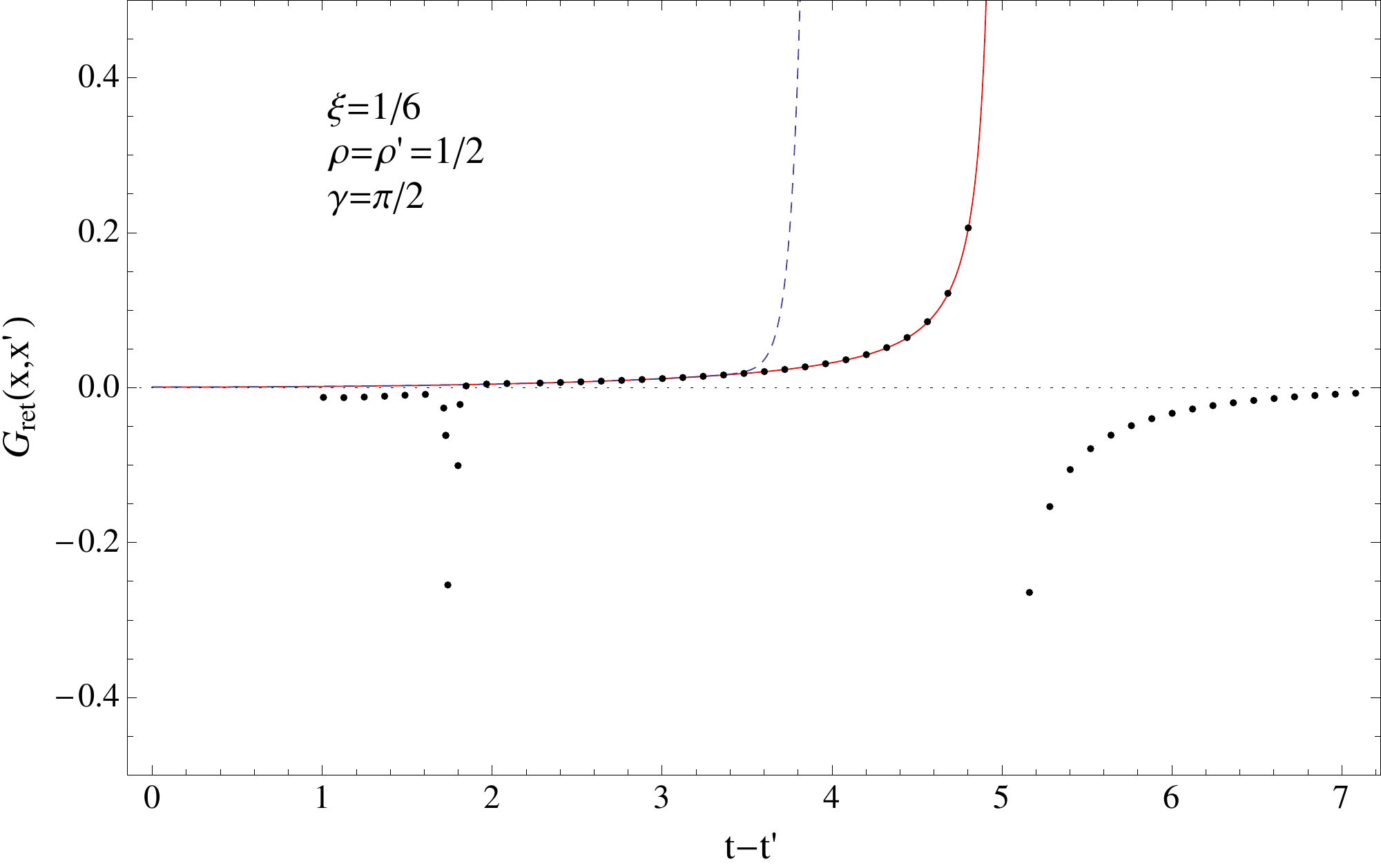}
 \end{center}
 \caption{\emph{Comparing Pad\'{e} approximant and Taylor series for $\theta(-\sigma(x,x')) V(x,x')$ to `exact' Green function from quasinormal mode sum in Nariai spacetime with curvature coupling $\xi=1/8$ and $\xi=1/6$.} The Pad\'e approximated $V(x,x')$ (red line) is in excellent agreement with the quasinormal mode Green function (black dots) up to the normal neighborhood boundary, $t_{NN}\approx6.56993$ (top) and $t_{NN}\approx4.9956$ (bottom).}
 \label{fig:padeCompareSeriesNariai}
\end{figure}

The Green function in Nariai spacetime is known to be given exactly by a quasinormal mode sum \cite{Casals:Dolan:Ottewill:Wardell:2009,Beyer:1999} at sufficiently late times. We can therefore use the Green function calculated from a quasinormal mode sum to determine the effectiveness of the Pad\'e resummation. Figure \ref{fig:padeCompareSeriesNariai} compares the quasinormal mode calculated Green function\footnote{
There are some caveats with how the quasinormal Green function was used. The \emph{fundamental mode} ($n=0$) Green function was used and a \emph{singularity time offset} applied as described in Ref.~\cite{Casals:Dolan:Ottewill:Wardell:2009}. In the case where two singularities are present, two sets of fundamental mode Green functions were used, each shifted by an appropriate singularity time offset and matched at an intermediate point.} with both the original Taylor series representation and the Pad\'e resummed series for $V(x,x')$ for a range of cases. We use the Pad\'e approximant $P_{30}^{30}$, computed from the $60^{th}$ order Taylor series. In each case, the series representation (blue dashed line) diverges near its radius of convergence, long before the normal neighborhood boundary is reached. The Pad\'e resummed series (red line), however, remains valid much further and closely matches the quasinormal mode Green function (black dots) up to the point where the normal neighborhood boundary is reached.

As was shown in Refs.~\cite{Kay:Radzikowski:Wald:1997,Casals:Dolan:Ottewill:Wardell:2009}, the Green function in Nariai spacetime is singular whenever the points are separated by a null geodesic. Furthermore, in Ref.~\cite{Casals:Dolan:Ottewill:Wardell:2009} we have derived the functional form of these singularities and shown that they follow a four-fold pattern: $\delta(\sigma)$, $1/\pi \sigma$, $-\delta(\sigma)$, $-1/\pi \sigma$, depending on the number of caustics the null geodesic has passed through (this was also previously shown by Ori \cite{Ori1}). Within the normal neighborhood (where the Hadamard parametrix, \eqref{eq:Hadamard}, is valid), the $\delta(\sigma)$ singularities (i.e. at \emph{exactly} the null geodesic times) will be given by the term involving $U(x,x')$. However, a times other than the \emph{exact} null geodesic times, the Green function will be given fully by $V(x,x')$. For this reason, we expect $V(x,x')$ to reflect the singularities of the Green function near the normal neighborhood boundary

The Pad\'e approximant attempts to model the singularity of the function $V(x,x')$ (which occurs at the null geodesic time) by representing it as a rational function, i.e. a ratio of two power series. By its nature, this will only faithfully reproduce singularities of integer order. In the Nariai case, however, the asymptotic form of the the singularities is known exactly near the singularity times, $t_c$ \cite{Casals:Dolan:Ottewill:Wardell:2009}. In cases where the points are separated by an angle $\gamma\in (0,\pi)$ (i.e. away from a caustic), the singularities are expected to have a $1/(t-t'-t_c)$ behavior and it is reasonable to expect the Pad\'e approximant to reproduce the singularity well. When the points are not separated in the angular direction (i.e. at a caustic), however, the singularities behave like $1/(t-t'-t_c)^{3/2}$ and we cannot reasonably expect the Pad\'e approximant to accurately reflect this singularity without including a large number of terms in the denominator.

\begin{figure}
  \begin{center}
  \includegraphics[width=6cm]{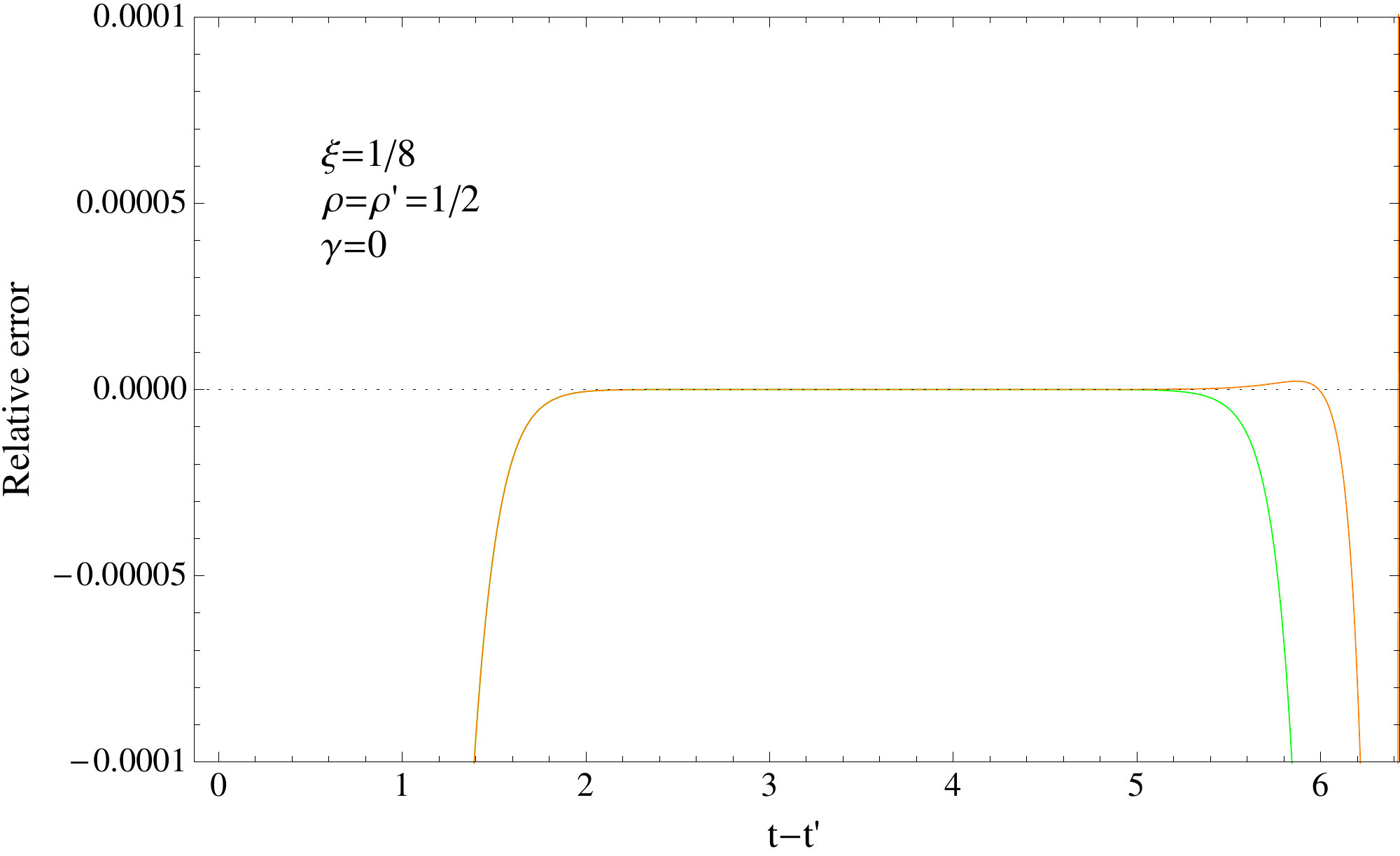}
 \end{center}
 \caption{\emph{Relative Error in Improved vs Regular Pad\'{e} Approximant.} The relative error in the improved Pad\'e approximant (orange) remains small closer to the singularity than the regular Pad\'e approximant (green).}
 \label{fig:padeSqrtNariai}
\end{figure}

Given knowledge of the functional form of the singularity, however, it is possible to improve the accuracy of the Pad\'e approximant further. For a singularity of the form $1/S(t)$, we first multiply the Taylor series by $S(t)$.  The result should then have either no singularity, or have a singularity which can be reasonably represented by a power series. The Pad\'e approximant of this new series is then calculated and the result is divided by $S(t)$ to give an \emph{improved Pad\'e approximant}. This yields an approximant which includes the exact form of the singularity and more closely matches the exact Green function near the singularity. In Fig.~\ref{fig:padeSqrtNariai}, we illustrate the improvement with an example case. We consider a static point in Nariai spacetime and compute the error in the Pad\'e approximant relative to the quasinormal mode Green function (with $n\le8$). The regular Pad\'e approximant is shown in green while the improved Pad\'e approximant is show in orange. The relative error remains small closer to the singularity for the improved Pad\'e approximant case than for the the standard Pad\'e approximant. Note that the error for early times arises from the failure of the quasinormal mode sum to converge and does not reflect error in the series approximations.

\subsection{Schwarzschild} \label{subsec:padeSchw}
\begin{figure}
  \begin{center}
  \includegraphics[width=6cm]{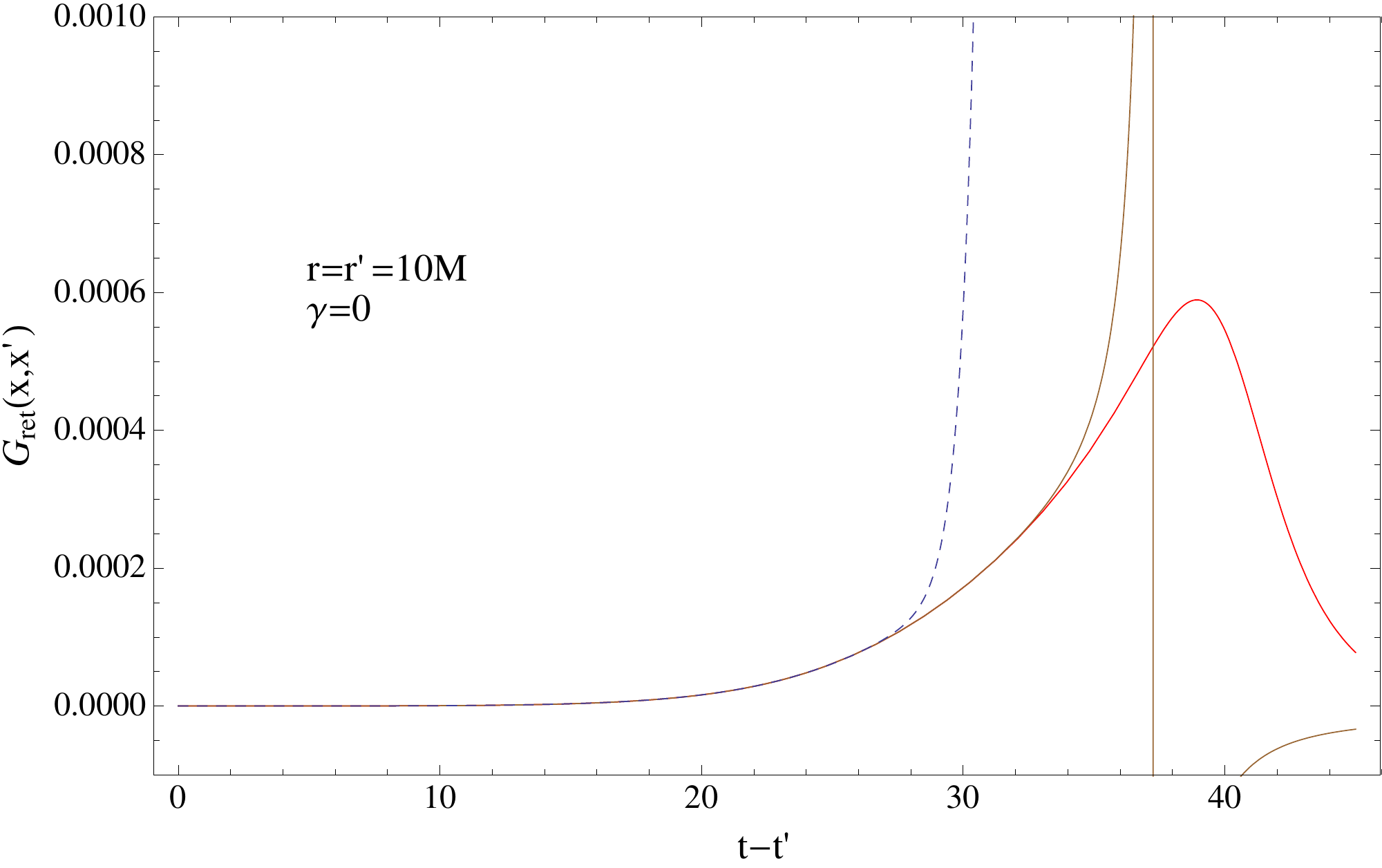}
 \end{center}
 \caption{\emph{Comparing Pad\'{e} to Taylor series for Schwarzschild} in the case of a static particle at r=10M. The Pad\'e approximants $P^{26}_{26}$ (red line) and $P^{24}_{26}$ (brown line) are likely to represent $V(x,x')$ more accurately near the normal neighborhood boundary (at $t-t'\approx46.2471M$) than the regular Taylor series (blue dashed line).}
 \label{fig:padeCompareSeriesSchw}
\end{figure}

For the Schwarzschild case, there is no quasinormal mode sum with which to compare the Pad\'e approximated series\footnote{A quasinormal mode sum could be computed for the Schwarzschild case, but would be augmented by a branch cut integral \cite{Casals:Dolan:Ottewill:Wardell:2009}. This calculation is in progress but has yet to be completed}. However, given the success in the Nariai case, we remain optimistic that Pad\'e approximation will be successful for Schwarzschild. In an effort to estimate the effectiveness of the Pad\'e approximant, we compare in Fig.~\ref{fig:padeCompareSeriesSchw} the series expression for $V(x,x')$ with two different Pad\'e resummations, $P_{26}^{24}$ and $P_{26}^{26}$.  The Pad\'e approximant extends the validity beyond the radius of convergence of the series, but is less successful at reaching the normal neighborhood boundary ($t-t'=t_{NN}\approx 46.2471M$) than in the Nariai case.

The failure of the Pad\'e approximant to reach the normal neighborhood boundary can be understood by the presence of extraneous singularities in the Pad\'e approximant. The zeros of the denominator, $S(t-t')=0$, give rise to singularities which occur at times earlier that the null geodesic time. It is possible that this problem could be reduced to a certain extent using the knowledge of the functional form of the singularities to compute an \emph{improved} Pad\'e approximant (as was successful in the Nariai case). However, to the authors knowledge, the structure of the singularities in Nariai spacetime is not yet known. While it may be possible to adapt the work of Ref.~\cite{Casals:Dolan:Ottewill:Wardell:2009} to find the asymptotic form of the singularities in Schwarzschild, without knowledge of the exact Green function we would not be able to dermine whether an improved Pad\'e approximant would truly give an improvement. We therefore leave such considerations for later work.

\subsection{Convergence of the Pad\'e Sequence} \label{subsec:Pade-convergence}
The use of Pad\'e approximants has shown remarkable success in improving the accuracy and domain of the series representation of $V(x,x')$. However, this improvement has not been quantified. There is no general way to determine whether the Pad\'e approximant is truly approximating the correct function, $V(x,x')$ or the domain in which it is valid \cite{NumericalRecipes}. In this subsection, we nonetheless attempt to gain some insight into the validity of the Pad\'e approximants.

The first issue to consider is the presence of extraneous poles in the Pad\'e approximants. In Sec.~\ref{subsec:padeNariai}, the Pad\'e approximant was unable to exactly represent the $1/(t-t'-t_c)^{3/2}$ singularity at $t_c \approx 6.12$ and instead represented it by three (real-valued) simple poles (at $t-t' \approx 6.522, 6.854 \text{ and } 9.488$). This leads to the Pad\'e approximant being a poor representation of the function near the poles. As was shown in Fig.~\ref{fig:padeSqrtNariai}, having exact knowledge of the singularity allows the calculation of an improved Pad\'e approximant without extraneous singularities\footnote{The improved Pad\'e approximant has zeros in its denominator at $t-t' \approx 2.64, 6.51, 6.59 \text{and} 8.73$. The apparently extraneous singularity within the normal neighborhood (at $t-t'\approx 2.64$) does not cause any difficulty as the numerator also goes to zero at this point.}.

With extraneous poles dealt with, we consider the convergence of the Pad\'e sequence of diagonal and sub-diagonal Pad\'e approximants, 
\begin{equation}
 P = \{P_0^0,~ P_1^0,~ P_1^1,~ P_2^1,~ P_2^2,~ P_3^2,~ P_3^3, \cdots \},
\end{equation}
with $P_N$ being the $N$-th element of the sequence. The convergence of the Pad\'e approximant sequence is determined by the behavior of the denominators, $S_N$ for large $N$ \cite{Bender:Orszag}. Provided $S(x,x')_N$ is not small, the Pad\'e sequence will converge quickly toward the actual value of $V(x,x')$. When the first root of the denominator is at the null geodesic time (i.e. the normal neighborhood boundary), we can, therefore, be optimistic that the Pad\'e sequence will remain convergent until this root is reached and the Pad\'e approximants will accurately represent the function.

To highlight the improvements made by using Pad\'e approximants over regular Taylor series, we introduce the Taylor sequence (i.e. the sequence of partial sums of the series), $T=\{T_0, T_1, T_2, \cdots\}$, with $N$-th element,
\begin{equation}
 T_N = \sum_{n=0}^{N/2} v_{n} (t-t')^{2n}.
\end{equation}
The two sequences $P_N$ and $T_N$ require approximately the same number of terms in the original Taylor series, so a direct comparison of their convergence will illustrate the improved convergence of the Pad\'e approximants.

\begin{figure}
  \begin{center}
  \includegraphics[width=6cm]{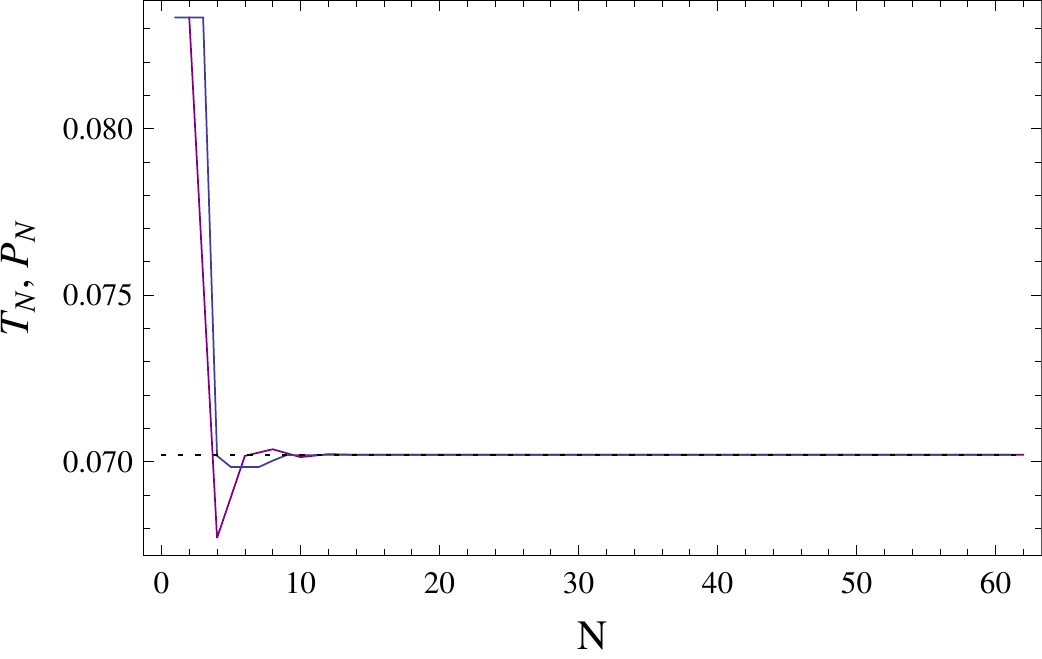}
  \includegraphics[width=6cm]{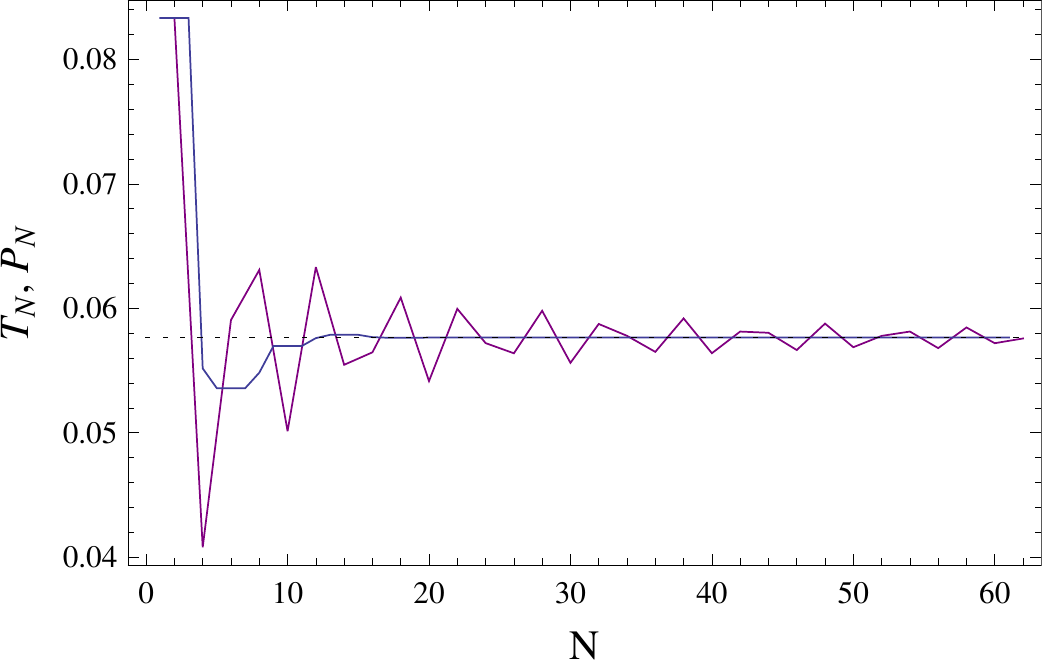}
  \includegraphics[width=6cm]{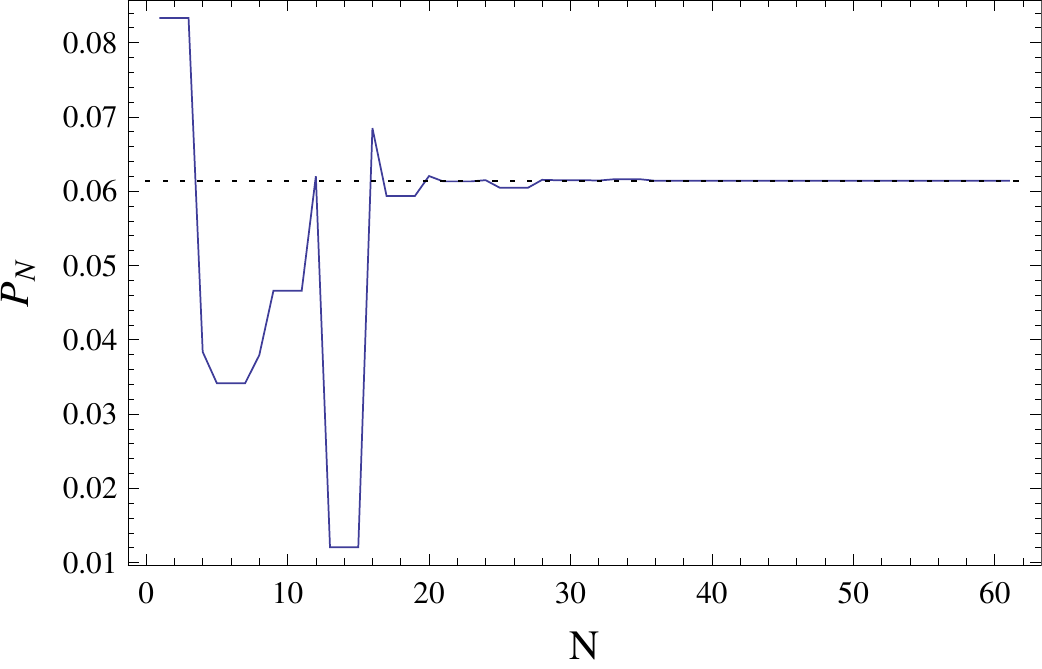}
  \includegraphics[width=6cm]{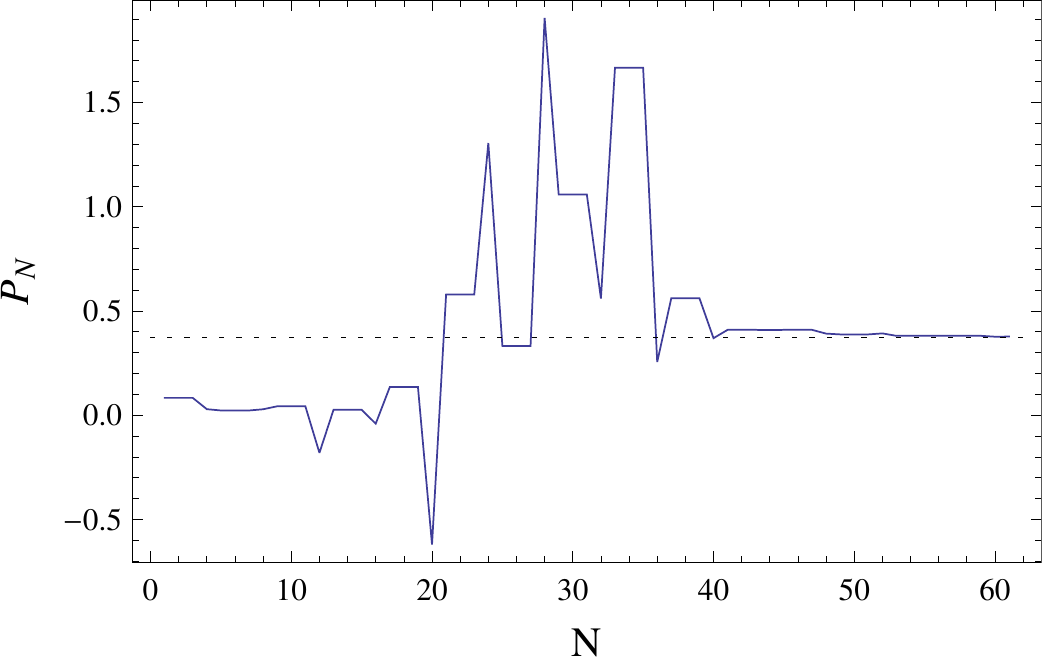}
 \end{center}
 \caption{\emph{Convergence of the Taylor and Pad\'{e} Sequences} for the case of static points at $\rho=1/2$ in the Nariai spacetime, with $\xi=1/8$. Within the radius of convergence of the Taylor series, both Pad\'e (blue line) and Taylor (purple line) sequences converge to the exact Green function (black dotted line) as calculated from a quasinormal mode sum with $n\le6$. The Pad\'e sequence converges faster, particularly at larger times. Outside the radius of convergence of the Taylor series, only the Pad\'e sequence is convergent. The top left plot is at a time $(t-t')=2$, top right plot is at $(t-t')=3.3$, bottom left at $(t-t')=5$, bottom right at $(t-t')=6.3$.}
 \label{fig:padeConvergenceNariai}
\end{figure}

In Fig.~\ref{fig:padeConvergenceNariai}, we plot the Pad\'e sequence (blue line) and Taylor sequence (purple line) for the case of static points at $\rho=1/2$ in the Nariai spacetime, with $\xi=1/8$. For early times (eg. $(t-t')=2$, it is clear that both Pad\'e and Taylor sequences converge very quickly. At somewhat later times (eg. $(t-t')=3.3$), both sequences appear to remain convergent, but the Pad\'e sequence is clearly converging much faster than the Taylor sequence. Outside the radius of convergence of the Taylor series (eg. $(t-t')=5, 6.3$), the Pad\'e sequence is slower to converge, but appears to still do so.

\begin{figure}
  \begin{center}
  \includegraphics[width=6cm]{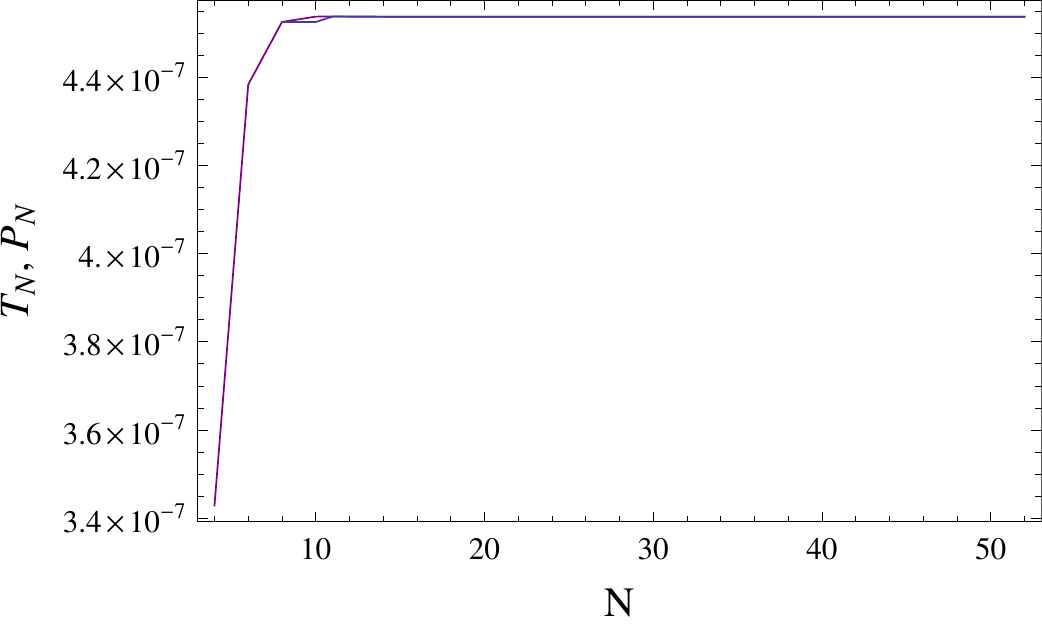}
  \includegraphics[width=6cm]{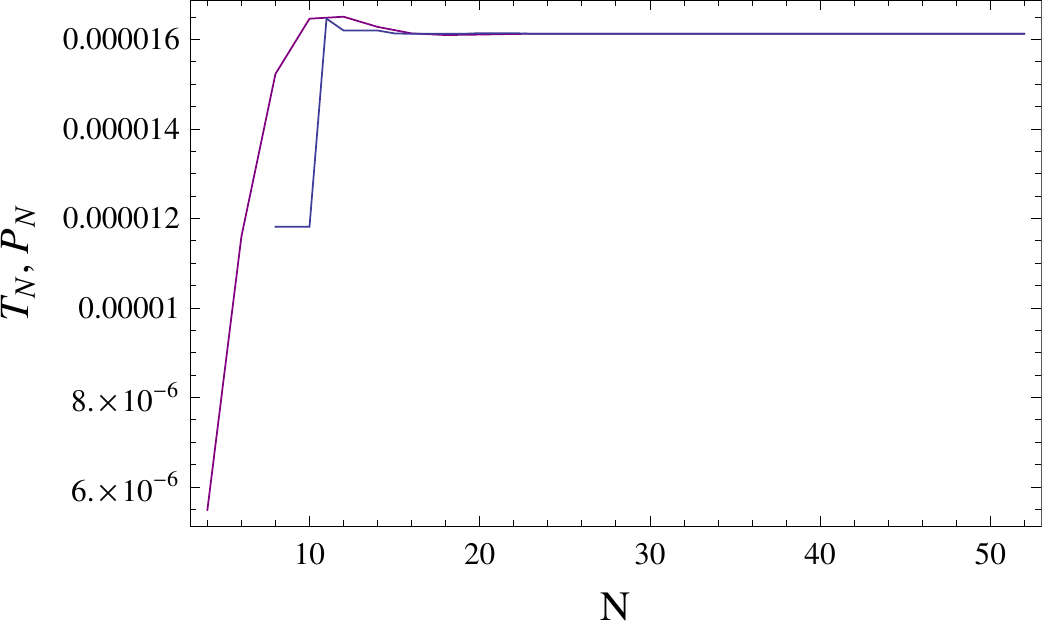}
  \includegraphics[width=6cm]{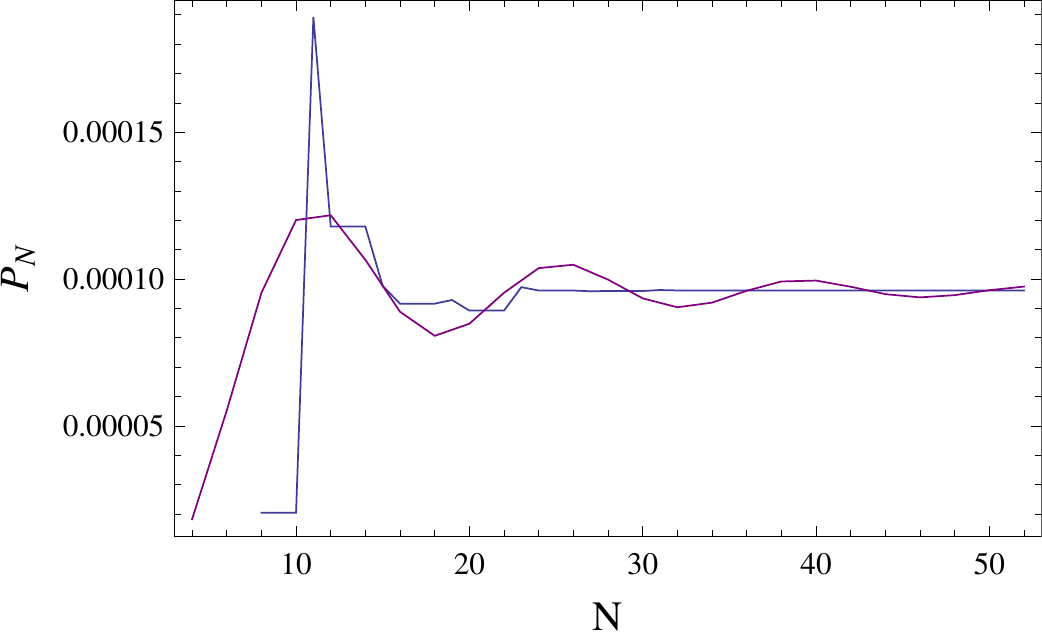}
  \includegraphics[width=6cm]{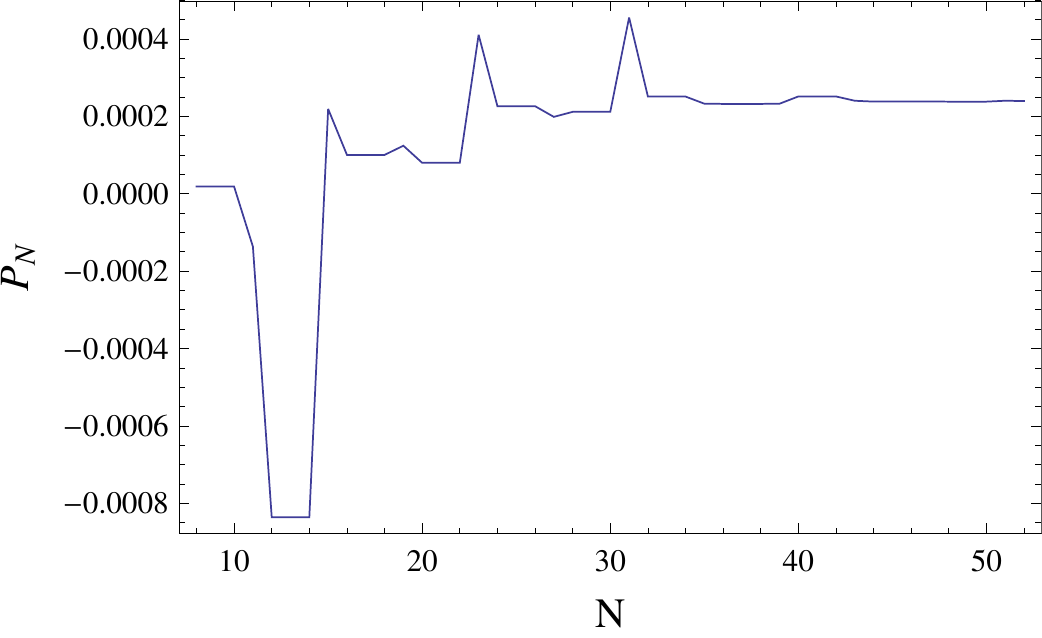}
 \end{center}
 \caption{\emph{Convergence of the Taylor and Pad\'{e} Sequences} for the case of static points at $r=10M$ in the Schwarzschild spacetime. The top left plot is at a time $(t-t')=10M$, top right plot is at $(t-t')=20M$, bottom left plot is at $(t-t')=27M$, bottom right plot is at $(t-t')=32M$.}
 \label{fig:padeConvergenceSchw}
\end{figure}

In Fig.~\ref{fig:padeConvergenceSchw}, we again plot the Pad\'e sequence, this time for the case of static points at $r=10M$ in the Schwarzschild spacetime. As in the Nariai case, for early times (eg. $(t-t')=10M$, it both Pad\'e and Taylor sequences are converging very quickly. At slightly later times (eg. $(t-t')=20M, 27M$) the convergence of the Pad\'e sequence is better than the Taylor sequence. Outside the radius of convergence of the Taylor series (eg. $(t-t')=32M$), the Pad\'e sequence is slower to converge, but appears to still do so. The convergence of the series is slower in the Schwarzschild case than in the Nariai case. This is an indication that using more terms may yield a better result\footnote{In the Nariai case (with $\xi=1/8$), the Taylor series for $V(x,x')$ starts at order $(t-t')^0$ and has been calculated to order $(t-t')^{60}$, while for Schwarzschild it starts at $(t-t')^4$ and has been calculated to order $(t-t')^{52}$. Since the Nariai series has several extra orders, it is reasonable to expect it to be a better approximation than the Schwarzschild series.}.

\section{Conclusions}

In this paper we have presented an extension of the Hadamard-WKB method of Anderson and Hu \cite{Anderson:2003} to the Nariai spacetime. We have also demonstrated the use of an alternative WKB method \cite{Winstanley:2007,Howard:1985}, which allows very high order terms in the Taylor series to be calculated efficiently (on a computer) for Schwarzschild, Nariai and other spherically symmetric spacetimes. This allowed the series expansion of $V(x,x')$ (appearing in the Hadamard parametrix of the Green function) to be computed to significantly higher order than was done previously in Refs.~\cite{Anderson:2003,Anderson:Eftekharzadeh:Hu:2006}. These high order expansions facilitated an investigation of the convergence properties of the series. We also demonstrated the huge benefit of Pad\'e approximants to improving the domain and convergence of the series.

This paper serves a dual purpose
\begin{enumerate}
 \item To discuss the calculation of the quasilocal Green function in Nariai spacetime, as required by Ref.~\cite{Casals:Dolan:Ottewill:Wardell:2009}.
\item To investigate the potential for applying the same techniques in the Schwarzschild spacetime, with the goal of computing an accurate quasilocal Green function for use in a matched expansion calculation of the self-force.
\end{enumerate}

We have found that, using Pad\'e approximants, it is possible compute the quasilocal Green function in Nariai spacetime to high accuracy to within a short distance of the normal neighborhood boundary. Even without the use of Pad\'e approximants, the Taylor approximated series gives good accuracy within a large part of the quasilocal region. This gives confidence in their use for matched expansion calculations in Ref.~\cite{Casals:Dolan:Ottewill:Wardell:2009}.

With regard to the Schwarzschild case, we find that the quasilocal calculation of the Green function is in good standing and should be usable in matched expansion calculations once techniques for computing the `distant-past' Green function have been fully developed. Both Pad\'e and Taylor sequences remain convergent within a large part of the normal neighborhood. The use of Pad\'e approximants is not quite as successful as for the Nariai case, but we remain optimistic that knowledge of the structure of the singularities in Schwarzschild may allow for the use of \emph{improved} Pad\'e approximants as discussed in Sec.~\ref{subsec:padeSchw}.

The orders of the series calculated for Nariai ($60$-th) and Schwarzschild ($52$-nd) were the maximum possible within a reasonable time ($\sim 1$ day on a modern Linux desktop). Although these are considerably high order series, one may still wonder whether they are sufficiently high for matched expansion calculations. It is clear from Sec.~\ref{subsec:Pade-convergence} that the higher order terms only have a significant contribution near the radius of convergence (for the Taylor series) or normal neighborhood boundary (for the Pad\'e approximant). As the Hadamard parametrix is only valid within the normal neighborhood, we consider the fact that the Pad\'e approximant is accurate to within a short distance of the normal neighborhood boundary to be confirmation that the series has been calculated to sufficiently high order (in particular for the Nariai case). Additionally, within the matching region used in Ref.~\cite{Casals:Dolan:Ottewill:Wardell:2009} for Nariai, we clearly have computed a sufficient number of coefficients to give the Green function to high accuracy. We can be optimistic that this is also the case for Schwarzschild: the quasilocal series is accurate long after the time when the quasinormal mode sum is expected to be convergent ($t-t'= 2 r_* \approx 12.77 M$, for the case of a static particle at $r=10M$ considered here).

Our analysis has remained focused primarily on the case of one dimensional series. This was done for reasons of simplicity and clarity. For multi-dimensional series, one could re-express each of the coordinates in terms of a single parameter, as was done in Sec.~\ref{sec:convergence} for the case of a circular geodesic in Schwarzschild. Alternatively, one could make use of the extension of the Pad\'e approximant to double and higher dimensional power series as developed by Chisholm \cite{Chisholm:1973,Chisholm:McEwan:1974}.

\section{Acknowledgments}
MC is grateful to the Department of Physics and Astronomy of the University of Mississippi for its hospitality during the preparation of this paper. MC was partially funded by Funda\c c\~ao para a Ci\^encia e Tecnologia (FCT) - Portugal through project PTDC/FIS/64175/2006. MC, BW and SD are supported by the Irish Research Council for Science, Engineering and Technology, funded by the National Development Plan.

\appendix

\end{document}